# Slowly fading super-luminous supernovae that are not pair-instability explosions

M. Nicholl[1], S. J. Smartt[1], A. Jerkstrand[1], C. Inserra[1], M. McCrum[1], R. Kotak[1], M. Fraser[1], D. Wright[1], T.-W. Chen[1], K. Smith[1], D. R. Young[1], S. A. Sim[1], S. Valenti[2,3], D. A. Howell[2,3], F. Bresolin[4], R. P. Kudritzki[4], J. L. Tonry[4], M. E. Huber[4], A. Rest[5], A. Pastorello[6], L. Tomasella[6], E. Cappellaro[6], S. Benetti[6], S. Mattila[7,8], E. Kankare[7,8], T. Kangas[8], G. Leloudas[9,10], J. Sollerman[11], F. Taddia[11], E. Berger[12], R. Chornock[12], G. Narayan[12], C. W. Stubbs[12], R. J. Foley[12], R. Lunnan[12], A. Soderberg[12], N. Sanders[12], D. Milisavljevic[12], R. Margutti[12], R. P. Kirshner[12,13], N. Elias-Rosa[14], A. Morales-Garoffolo[14], S. Taubenberger[15], M. T. Botticella[16], S. Gezari[17], Y. Urata[18], S. Rodney[19], A. G. Riess[19], D. Scolnic[19], W. M. Wood-Vasey[20], W. S. Burgett[4], K. Chambers[4], H. A. Flewelling[4], E. A. Magnier[4], N. Kaiser[4], N. Metcalfe[21], J. Morgan[4], P. A. Price[22], W. Sweeney[4] & C. Waters[4]

[1]Astrophysics Research Centre, School of Mathematics and Physics, Queen's University Belfast, Belfast BT7 1NN, UK.

[2]Las Cumbres Observatory Global Telescope Network, 6740 Cortona Drive, Suite 102 Goleta, California 93117, USA.

[3]Department of Physics, Broida Hall, University of California, Santa Barbara, California 93106, USA.

[4]Institute of Astronomy, University of Hawaii, 2680 Woodlawn Drive, Honolulu, Hawaii 96822, USA.

[5]Space Telescope Science Institute, 3700 San Martin Drive, Baltimore, Maryland 21218, USA.

[6]INAF, Osservatorio Astronomico di Padova, Vicolo dell'Osservatorio 5, 35122 Padova, Italy.

[7]Finnish Centre for Astronomy with ESO (FINCA), University of Turku, Väisäläntie 20, FI-21500 Piikkiö, Finland.

[8]Tuorla Observatory, Department of Physics and Astronomy, University of Turku, Väisäläntie 20, FI-21500 Piikkiö, Finland.

[9]The Oskar Klein Centre, Department of Physics, Stockholm University, 10691 Stockholm, Sweden.

[10]Dark Cosmology Centre, Niels Bohr Institute, University of Copenhagen, 2100 Copenhagen, Denmark.

[11]The Oskar Klein Centre, Department of Astronomy, Stockholm University, 10691 Stockholm, Sweden.

[12]Harvard-Smithsonian Center for Astrophysics, 60 Garden Street, Cambridge, Massachusetts 02138, USA.

[13]Kavli Institute for Theoretical Physics, University of California Santa Barbara, Santa Barbara, California 93106, USA.

[14]Institut de Ciències de l'Espai (IEEC-CSIC), Facultat de Ciències, Campus UAB, 08193 Bellaterra, Spain.

[15]Max-Planck-Institut für Astrophysik, Karl-Schwarzschild-Strasse 1, 85741 Garching, Germany.

[16]INAF–Osservatorio Astronomico di Capodimonte, Salita Moiariello 16, I-80131 Napoli, Italy.

[17]Department of Astronomy, University of Maryland, College Park, Maryland 20742-2421, USA.

[18]Institute of Astronomy, National Central University, Chung-Li 32054, Taiwan.

[19]Department of Physics and Astronomy, Johns Hopkins University, 3400 North Charles Street, Baltimore, Maryland 21218, USA.

[20]Pittsburgh Particle Physics, Astrophysics, and Cosmology Center, Department of Physics and Astronomy, University of Pittsburgh, 3941 O'Hara Street, Pittsburgh, Pennsylvania 15260, USA.

[21]Department of Physics, Durham University, South Road, Durham DH1 3LE, UK.

[22]Department of Astrophysical Sciences, Princeton University, Princeton, New Jersey 08544, USA.

**Super-luminous supernovae[1–4] that radiate more than $10^{44}$ ergs per second at their peak luminosity have recently been discovered in faint galaxies at redshifts of 0.1–4. Some evolve slowly, resembling models of 'pair-instability' supernovae[5,6]. Such models involve stars with original masses 140–260 times that of the Sun that now have carbon–oxygen cores of 65–130 solar masses. In these stars, the photons that prevent gravitational collapse are converted to electron–positron pairs, causing rapid contraction and thermonuclear explosions. Many solar masses of $^{56}$Ni are synthesized; this isotope decays to $^{56}$Fe via $^{56}$Co, powering bright light curves[7,8]. Such massive progenitors are expected to have formed from metal-poor gas in the early Universe[9]. Recently, supernova 2007bi in a galaxy at redshift 0.127 (about 12 billion years after the Big Bang) with a metallicity one-third that of the Sun was observed to look like a fading pair-instability supernova[1,10]. Here we report observations of two slow-to-fade super-luminous supernovae that show relatively fast rise times and blue colours, which are incompatible with pair-instability models. Their late-time light-curve and spectral similarities to supernova 2007bi call the nature of that event into question. Our early spectra closely resemble typical fast-declining super-luminous supernovae[2,11,12], which are not powered by radioactivity. Modelling our observations with 10–16 solar masses of magnetar-energized[13,14] ejecta demonstrates the possibility of a common explosion mechanism. The lack of unambiguous nearby pair-instability events suggests that their local rate of occurrence is less than $6 \times 10^{-6}$ times that of the core-collapse rate.**

The discovery of a luminous transient, PTF12dam, was first reported[15] by the Palomar Transient Factory on 23 May 2012. We recovered the transient in Pan-STARRS1 (Panoramic Survey Telescope and Rapid Response System) 3π survey data, between 13 and 29 April 2012, at right ascension (RA) 14 h 24 min 46.21 s and declination (dec.) +46° 13′ 48.66″. We triggered spectroscopic follow-up, beginning with Gran Telescopio Canarias and the William Herschel Telescope (23–25 May 2012). No traces of hydrogen or helium were visible, leading to a type Ic classification, and strong host galaxy lines provided a redshift measurement $z = 0.107$ (ref. 15). A second, similar transient, PS1-11ap, was discovered in the Pan-STARRS1 Medium Deep Survey on 2 January 2011 (RA 10 h 48 min 27.72 s, dec. +57° 09′ 09.2″). Early spectra showed host galaxy emission lines at $z = 0.523$ (for details of the data, see Supplementary Information sections 1–3).

The high luminosity and slow decline of their light curves (Fig. 1, Extended Data Tables 1–3, Extended Data Fig. 1) marked out PTF12dam and PS1-11ap as potential SN 2007bi-like events: that is, they could be pair-instability supernova (PISN) candidates discovered soon after explosion. SN 2007bi was discovered well after maximum light. Although the peak was recovered in the R band[1], the light-curve rise and early spectra were missed. Because of the long diffusion timescale associated with the very massive ejecta in PISN models, the time to reach maximum light (≳100 days) is a crucial observational test. The rise time for SN 2007bi was estimated at 77 days (ref. 1), but this was based on a parabolic fit to the data around the peak, and so was not well constrained. Our Pan-STARRS1 images reveal multiple early detections of PTF12dam and PS1-11ap in $g_{P1}$, $r_{P1}$ and $i_{P1}$ bands at around 50 and 35 rest-frame days before peak brightness, respectively (Extended Data Fig. 2). PTF12dam is not detected in $z_{P1}$ images on 1 January 2012, 132 days before the peak. Although their light curves match the declining phases of SN 2007bi and the PISN models quite well, PTF12dam and PS1-11ap rise to maximum light a factor of ~2 faster than these models.

The spectra of PTF12dam and PS1-11ap show them to be similar supernovae. After 50 days from the respective light curve peaks, these spectra are almost identical to that of SN 2007bi at the same epoch (Fig. 2, Extended Data Table 4, Extended Data Fig. 3). The blue colours are in stark contrast to the predictions of PISN models[7,8] (Fig. 3, Extended Data Fig. 4), which show much cooler continua below 5,000 Å and marked drop-offs in the ultraviolet. Particularly around and after maximum light, PISN colours are expected to evolve to the red owing to increasing blanketing by iron group elements[7,8] abundant in their ejecta. We see no evidence of line blanketing in our spectra, even down to 2,000 Å (rest frame) in PS1-11ap, which suggests lower iron group abundances and a higher degree of ionization than in PISN models. Such conditions are fulfilled in models of ejecta reheated by magnetars—highly magnetic, rapidly rotating nascent pulsars[13,16,17]. The pressure of the magnetar wind on the inner ejecta can form a dense shell[13,14,17] at near-constant photospheric velocity. For PTF12dam, the velocities of spectral lines are close to 10,000 km s$^{-1}$ at all times. Intriguingly, the early spectra of our objects are very similar to those of superluminous supernovae of type I (refs 2, 11, 12) and evolve in the same way, but on longer timescales and with lower line velocities (Fig. 2).

Nebular modelling of SN 2007bi spectra has been used to argue[1] for large ejected oxygen and magnesium masses of 8–15$M_\odot$ and 0.07–0.13$M_\odot$, respectively (where $M_\odot$ is the solar mass). Such masses are actually closer to values in massive core-collapse models[18] than in PISN models, which eject ~40$M_\odot$ oxygen and ~4$M_\odot$ magnesium[1,8,9]. In the work reported in ref. 1, an additional 37$M_\odot$ in total of Ne, Si, S, and Ar were added to the model, providing a total ejecta mass consistent with a PISN. However, this was not directly measured[1], because these elements lack any identified lines. These constraints are important, so we investigated line formation in this phase using our own non-local thermodynamic equilibrium code[19] (Extended Data Fig. 5; Supplementary Information section 4). We found that the luminosities of [O I] 6,300, 6,364 Å, O I 7,774 Å and Mg I] 4,571 Å, and the feature at 5,200 Å ([Fe II] + Mg I), can be reproduced with 10–20$M_\odot$ of oxygen-dominated ejecta, containing ~0.001–1$M_\odot$ of iron, given reasonable physical conditions (singly ionized ejecta at a few thousand degrees). Thus, although the nebular modelling of SN 2007bi in ref. 1 provided a self-consistent solution for PISN ejecta, our calculations indicate that this solution is not unique, and has not ruled out lower-mass ejecta on the core-collapse scale (10$M_\odot$). Moreover, if the line at 5,200 Å is [Fe II], then both our model and the model of ref. 1 predict a dominant [Fe II] 7,155 Å line (at the low temperatures and high iron mass expected in PISN), which is not present in the observed spectra. To estimate the nickel mass needed to power PTF12dam radioactively, we constructed a bolometric light curve from our near-ultraviolet to near-infrared photometry (Fig. 4). PTF12dam is brighter than SN 2007bi, and fitting it with radioactively powered diffusion models[18,20] requires ~15$M_\odot$ of $^{56}$Ni in ~15–50$M_\odot$ of ejecta — combinations that are not produced in any physical model (Extended Data Fig. 6; Supplementary Information section 5.1). Furthermore, such large nickel fractions are clearly not supported by our spectra.

The combination of relatively fast rises and blue spectra, lacking ultraviolet line blanketing, shows that PTF12dam, PS1-11ap and probably SN 2007bi are not pair-instability explosions. We suggest here one model that can consistently explain the data. A magnetar-powered supernova can produce a light curve with the observed rise and decline rates as the neutron star spins down and reheats the ejecta[13,14,16,17]. It has been suggested that ~10% of core-collapses may form magnetars[14]. Although their initial-spin distribution is unknown, periods ≳1 ms are physically plausible. This mechanism has already been proposed for SN 2007bi[14], as well as for fast-declining superluminous type-I supernovae[2,21]. We fitted a magnetar-powered diffusion model[21,22] to the bolometric light curve of PTF12dam (Fig. 4), and found a good fit for magnetic field $B \approx 10^{14}$ G and spin period $P \approx 2.6$ ms, with an ejecta mass of ~10–16$M_\odot$. At peak, the r-band luminosities of PTF12dam and PS1-11ap are ~1.5 times that of SN 2007bi. Scaling our light curve by this factor, our model implies a similar ejected mass for SN 2007bi, with a slower-spinning magnetar ($P \approx 3.3$ ms), comparable to previous models[14]. If the magnetar theory is correct for normal superluminous type-I supernovae[2,21], our objects could be explained as a subset in which larger ejected masses and weaker magnetic fields result in slower photometric and spectroscopic evolution.

This leaves no unambiguous PISN candidates within redshift $z < 2$ (although possible examples exist at higher redshift[4]). We used the properties of the Pan-STARRS1 Medium Deep Survey (PS1 MDS, with a nightly detection limit of ~23.5 mag in g,r,i-like filters[21,23,24]) to constrain the local rate of stripped-envelope PISNs. We simulated PS1 MDS observations of 80, 100 and 130$M_\odot$ helium core PISN models[7] using our own Monte Carlo code[25] (Supplementary Information section 6), requiring an apparent magnitude <21 in at least one bandpass and a continuous 100-day (observer-frame) window of PS1 monitoring before considering an event a candidate PISN detection. Initially assuming a rate of $10^{-5}$ $R_{CCSN}$ (where $R_{CCSN}$ is the rate of occurrence of core-collapse supernovae[26]) for each model, we typically find five 100$M_\odot$ PISN candidates per year, at $z < 0.6$. The 130$M_\odot$ explosions have peak near-ultraviolet magnitudes of −22, resulting in apparent $r_{P1}$ and $i_{P1}$ magnitudes <20. PS1 should detect >90% of these within $z < 0.6$ (ten or more per year). Taking the 100$M_\odot$ result, the fact that we have not detected a single transient with these properties in the three years of PS1 is inconsistent with our assumed explosion rate at a level of 3.9$\sigma$ (Poisson statistics). This implies a 3$\sigma$ upper limit on their rate (within $z < 0.6$) of <6×$10^{-6}$ $R_{CCSN}$; even allowing another factor of ~2 to conservatively cover detection issues such as bad pixels or bright nearby stars, the rate of occurrence of super-luminous PISNs of type Ic must be at least a factor of ten lower than the overall rate of type-I superluminous supernovae[12]. PS1-11ap was our best candidate for a PISN explosion, but it fails to match the models. However, our calculation suggests that almost all the lower-mass (80$M_\odot$) PISNs would escape detection. Future searches for PISN candidates should target these fainter explosions at lower redshift (and larger volumes), or the more luminous candidates at $z > 1$.

We conclude that the classification of some slow-fading super-luminous supernovae[12] as radioactively driven is not supported observationally, and propose that these events can be united with virtually all known type-Ic super-luminous supernovae into a single class. Magnetar-powered models can explain their brightness and colours, and account for their diversity. The low upper limit we find for the rate of very massive PISNs reduces their potential impact on cosmic chemical evolution within $z \lesssim 1$. This relieves possible tension between their proposed existence in the nearby Universe, and the lack of detected chemical enrichment signatures in metal-poor stars and damped Lyman-α systems[27].

**Acknowledgements** We thank D. Kasen and L. Dessart for sending us their model data. The Pan-STARRS1 Surveys (PS1) have been made possible through contributions of the Institute for Astronomy, the University of Hawaii, the Pan-STARRS Project Office, the Max Planck Society (and its participating institutes, the Max Planck Institute for Astronomy, Heidelberg, and the Max Planck Institute for Extraterrestrial Physics, Garching), The Johns Hopkins University, Durham University, the University of Edinburgh, Queen's University Belfast, the Harvard-Smithsonian Center for Astrophysics, the Las Cumbres Observatory Global Telescope Network Incorporated, the National Central University of Taiwan, the Space Telescope Science Institute, NASA grant no. NNX08AR22G issued through the Planetary Science Division of the NASA Science Mission Directorate, National Science Foundation grant no. AST-1238877, and the University of Maryland. S.J.S. acknowledges FP7/2007-2013/ERC Grant agreement no. [291222]; J.L.T. and R. P. Kirshner acknowledge NSF grants AST-1009749, AST-121196; G.L. acknowledges Swedish Research Council grant no. 623-2011-7117; A.P., L.T., E.C., S.B. and M.T.B. acknowledge PRIN-INAF 2011. Work is based on observations made with the following telescopes: the William Herschel Telescope, Gran Telescopio Canarias, the Nordic Optical Telescope, Telescopio Nazionale Galileo, the Liverpool Telescope, the Gemini Observatory, the Faulkes North Telescope, the Asiago Copernico Telescope and the United Kingdom Infrared Telescope.

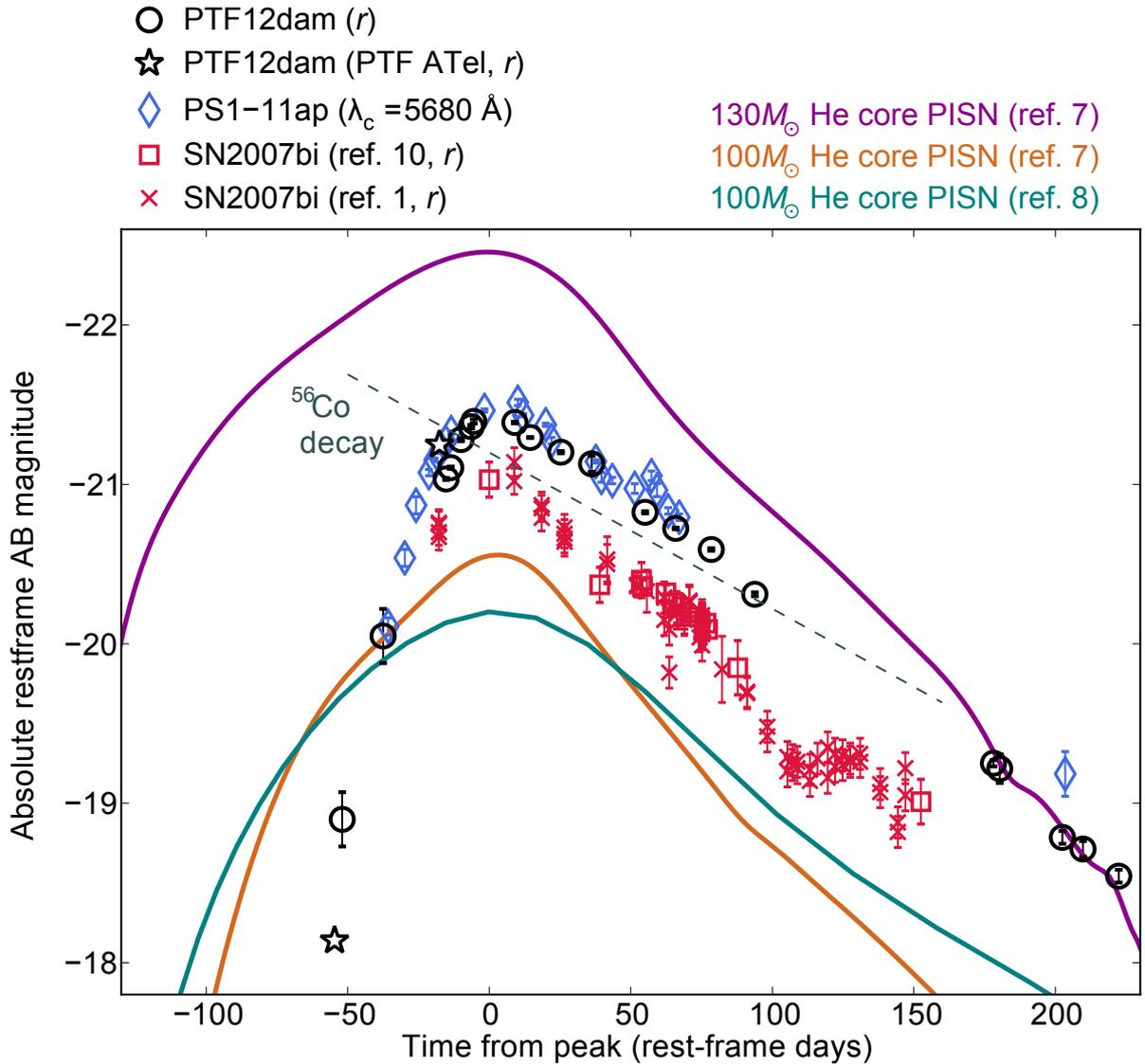

**Figure 1 | Optical light curves of slow-fading super-luminous supernovae.** Data for PTF12dam (including discovery data announced[15] by PTF) and SN 2007bi (from refs 1 & 10) are given in the SDSS (Sloan Digital Sky Survey) r band (central wavelength $\lambda_c = 6{,}230$ Å), while for PS1-11ap at $z = 0.523$, the PS1 $z_{P1}$ filter corresponds to a rest-frame filter of $\lambda_c = 5{,}680$ Å (width of passband $\lambda_{width} \approx 1{,}350$ Å), similar to SDSS r. The first three PS1-11ap points were transformed from $i_{P1}$ using the observed $i_{P1} - z_{P1}$ colour (see Supplementary Information sections 2 and 3 for details of the data, including k-corrections, colour transformations and extinction). The three supernovae (open symbols) display the same slow decline from maximum, matching the rate expected from $^{56}$Co decay (dashed line) with close to full γ-ray trapping (although similar declines can be generated for ~100 days after peak from magnetar spin-down[21]). Powering these high luminosities radioactively requires at least 3–7$M_\odot$

of $^{56}$Ni (refs 1, 10, 18 and 20), suggesting an extremely massive progenitor and possible pair-instability explosion[1]. Also shown are synthetic SDSS r-band light curves (solid lines) generated from published one-dimensional models[7,8] of PISNs from 100–130$M_\odot$ stripped helium cores. These fit the decline phase well, but do not match our early observations. The rise time of a PISN is necessarily long (rising 2.5 mag to peak in 95–130 days), because heating from $^{56}$Ni/$^{56}$Co decay occurs in the inner regions, and the resultant radiation must then diffuse through the outer ejecta, which typically has mass >80$M_\odot$ (ref. 7). Models with higher-dimensional outward mixing of $^{56}$Ni are likely to show even shallower gradients in the rising phase, while as-yet unexplored parameters such as rotation and magnetic fields will have little effect on the diffusion timescale, which is set by the mass, kinetic energy and opacity of the ejecta (see Supplementary Information section 5.2). The pre-peak photometry of PTF12dam and PS1-11ap show only a moderately slow rise over 50–60 days, which is therefore physically inconsistent with the PISN models. Error bars, ±1$\sigma$.

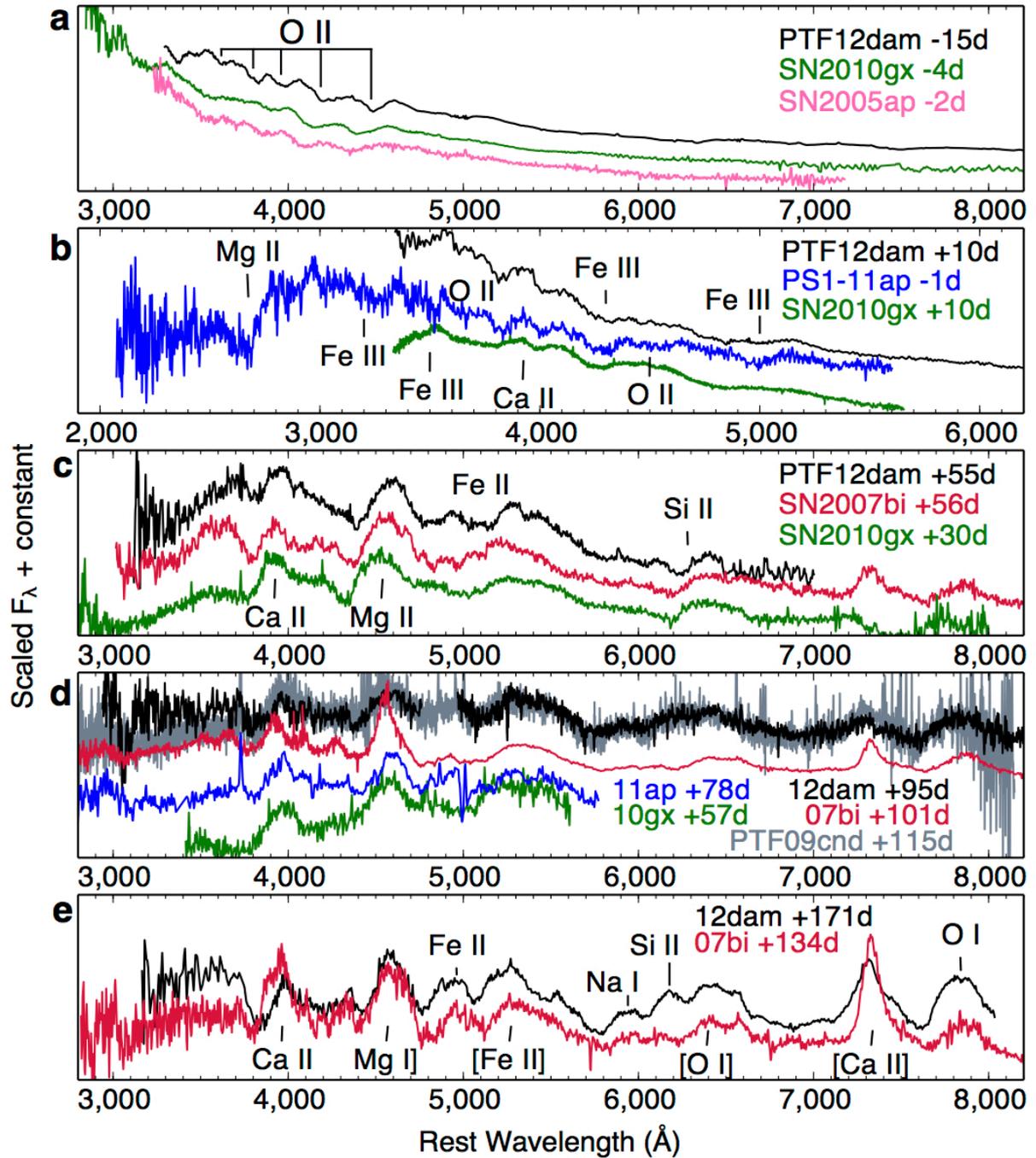

**Figure 2 | Spectral evolution of PTF12dam and PS1-11ap from superluminous supernovae of type I to SN 2007bi-like. a–e**, We show spectra of PTF12dam, PS1-11ap, SN 2007bi, and the well-studied superluminous supernovae of type I, SN 2010gx[11], SN 2005ap[2] and PTF09cnd[2]. Our spectra have been corrected for extinction and shifted to respective rest frames (details of reduction and analysis, including construction of model host continua for subtraction from **d** and **e**, in Supplementary Information section 3), and scaled to facilitate comparison. Phases are given in rest-frame days relative to maximum light. No hydrogen or helium are detected at any stage (near-infrared spectra of PTF12dam, obtained at +13 days and

+27 days, also show no He I; see Supplementary Information section 3). **a**, Before and around peak, our objects show the characteristic blue continua and O II absorptions common to super-luminous supernovae of type I/Ic[2,12,21], although the lines in the slowly evolving objects are at lower velocities than are typically seen in those events. **b,** Shortly after peak, Fe III features emerge, along with the Mg II and Ca II lines that dominate superluminous type I supernovae at this phase. **c**, By 55 days after peak, PTF12dam is almost identical to SN 2007bi. We note that these objects still closely resemble SN 2010gx, but seem to be evolving on longer timescales (consistent with the slower light-curve evolution). **d**, At ~100 days, PTF12dam also matches PTF09cnd[2], which faded slowly for a superluminous type I supernova after a 50-day rise. **e**, The spectra are now quasi-nebular, dominated by emission lines of Ca II H and K, Mg I] 4,571 Å, Mg I 5,183 Å + [Fe II 5,200] Å blend, [O I] 6,300, 6,364 Å, [Ca II] 7,291, 7,323 Å, and O I 7,774 Å, but some continuum flux is still visible. We find that the emission line intensities can be reproduced by ejecta from a $15M_\odot$ type I supernova at a few thousand degrees, without requiring a large mass of iron (Supplementary Information section 4).

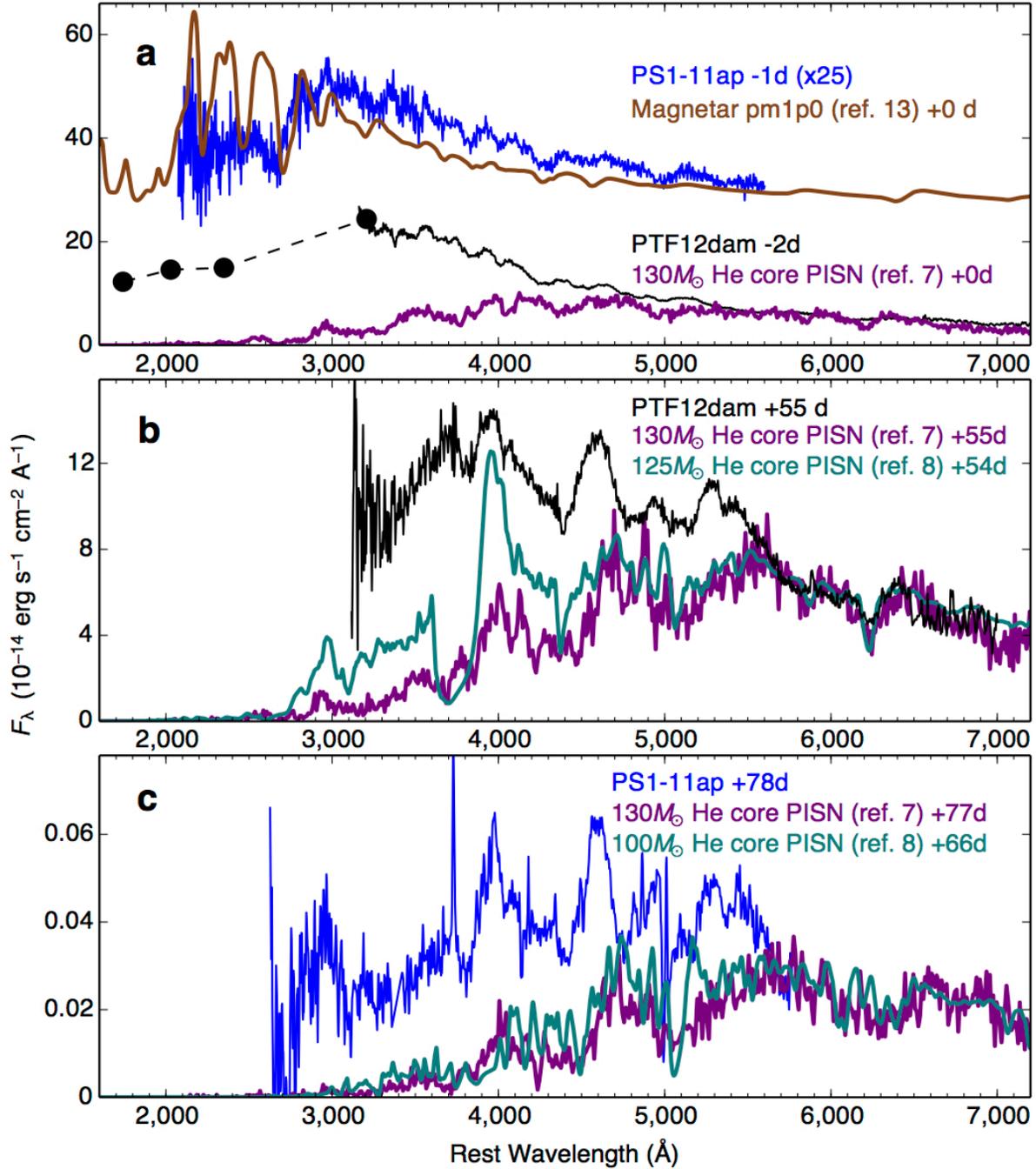

**Figure 3 | Spectral comparison with pair-instability and magnetar-driven supernova models. a-c,** We compare our ultraviolet and optical data to the predictions of PISN[7,8,13] and magnetar models[13] (lines in models are identified in refs 8 and 13). The absence of narrow lines and hydrogen/helium seems to make interaction-powered colliding-shell models unlikely (for example, the pulsational pair-instability; see Supplementary Information section 5.3). Model spectra are matched to the observed flux in the region 5,500–7,000 Å. **a,** We compare PS1-11ap to a Wolf–Rayet progenitor magnetar model (pm1p0[13]) at peak light (model spectra at later epochs do not currently exist in the literature). The magnetar energy input is equivalent

to several solar masses of $^{56}$Ni, in ejecta of only 6.94$M_\odot$. The high internal-energy-to-ejecta-mass ratio keeps the ejecta hot and relatively highly ionized, resulting in a blue continuum to match our observations. Moreover, this energy source does not demand the high mass of metals intrinsic to the PISN scenario[7,8]. Redward of the Mg II line at 2,800 Å, this model shows many of the same Fe III and O II lines dominating the observed spectra, although the strengths of the predicted Si III and C III lines in the near-ultraviolet are greater than those observed in PS1-11ap. We also compare PTF12dam at peak to a 130$M_\odot$ He core PISN model[7]. The model spectrum has intrinsically red colours below 5,000 Å due to many overlapping lines from the large mass of iron-group elements and intermediate-mass elements. Our rest-frame ultraviolet spectra of PS1-11ap, and ultraviolet photometry of PTF12dam, show that the expected line blanketing/absorption is not observed. **b,** PTF12dam compared to models of 125–130$M_\odot$ PISNs[7,8] at 55 days. Although the observed spectrum has cooled, the models still greatly under-predict the flux blueward of 5,000 Å. **c,** PS1-11ap, at 78 days, compared to 100–130$M_\odot$ PISN models[7,8] at similar epochs. Again, our observations are much bluer than PISN models. In particular, PS1-11ap probes the flux below 3,000 Å, where we see the greatest discrepancy.

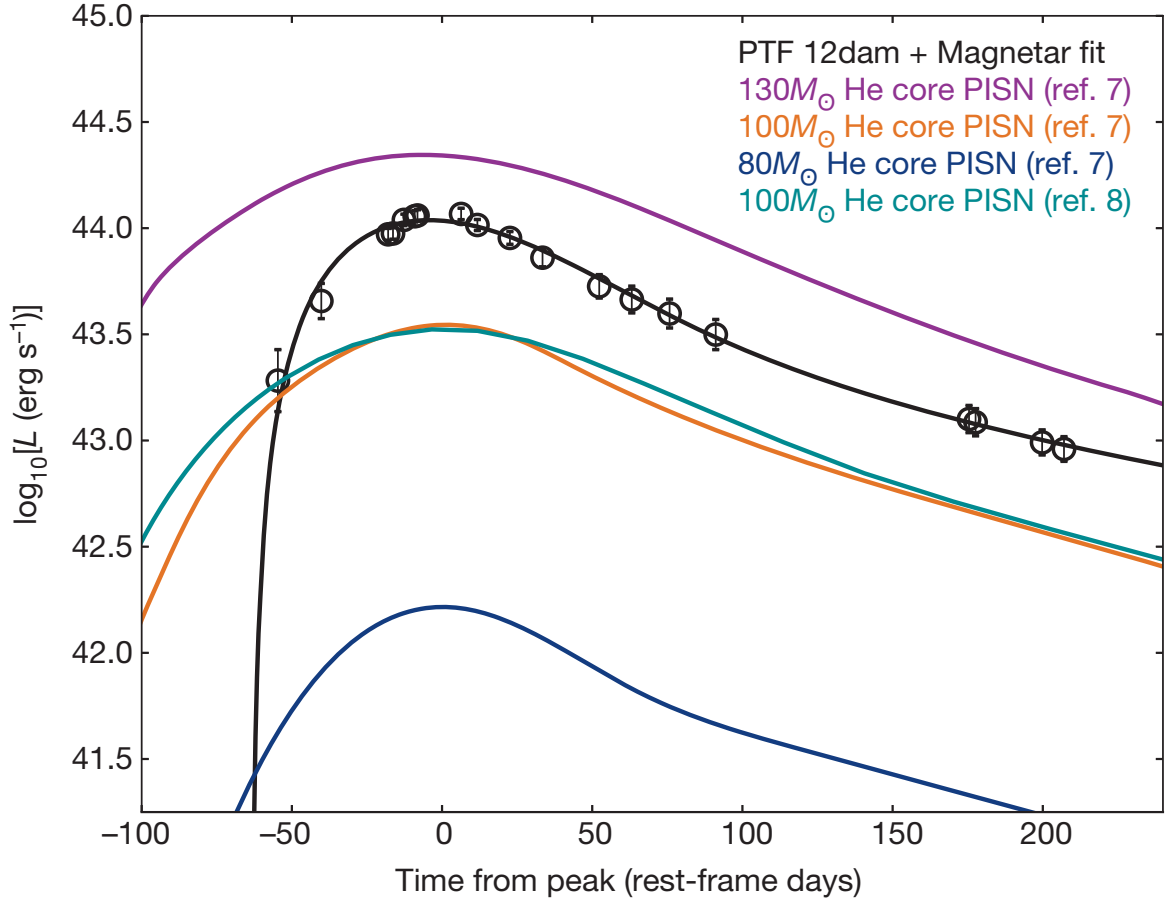

**Figure 4 | Bolometric light curve and magnetar fit.** Our PTF12dam bolometric light curve (open circles), comprising Swift observations in the near-ultraviolet, extensive *griz* imaging, and multi-epoch near-infrared (*JHK*) data (Supplementary Information section 5), is well fitted by our semi-analytic magnetar model[21] (black line) (see Supplementary Information section 5.4). This model, with magnetic field $B \approx 10^{14}$ G and spin period $P \approx 2.6$ ms, can fit both the rise and decay times of the light curve. A large ejecta mass of ~10–16$M_\odot$ is required—significantly higher than typically found for type Ibc supernovae[28], but similar to the highest estimates for SN 2011bm[29] and SN 2003lw[30] (though well below the >80$M_\odot$ expected in PISNs). In the context of the magnetar model, the parameters of our fit are consistent with the observed spectroscopic relation to super-luminous supernovae of type I. Fits to a sample of such objects using the same model[21] found uniformly lower ejected masses and higher magnetic fields than in PTF12dam. The large ejecta mass here results in a slow light-curve rise and broad peak compared to other super-luminous supernovae of type Ic[2,3,21], and would explain the slower spectroscopic evolution, including why the spectrum is not fully nebular at 200 days. The weaker *B* field means that the magnetar radiates away its rotational energy less

rapidly, so that more of the heating takes place at later times; this gives the impression of a radioactive tail. Higher ejected mass and weaker magnetar wind may account for the lower velocities in slowly declining events. Also shown for comparison are bolometric light curves of model PISNs[7,8] from 80–130$M_\odot$ He cores (coloured lines). Although PISNs from less massive progenitors do show faster rise times, the rise of PTF12dam is too steep to be consistent with the PISN explosion of a He core that is sufficiently massive to generate its observed luminosity. Errors bars, ±1$\sigma$ photometry, combined in quadrature.

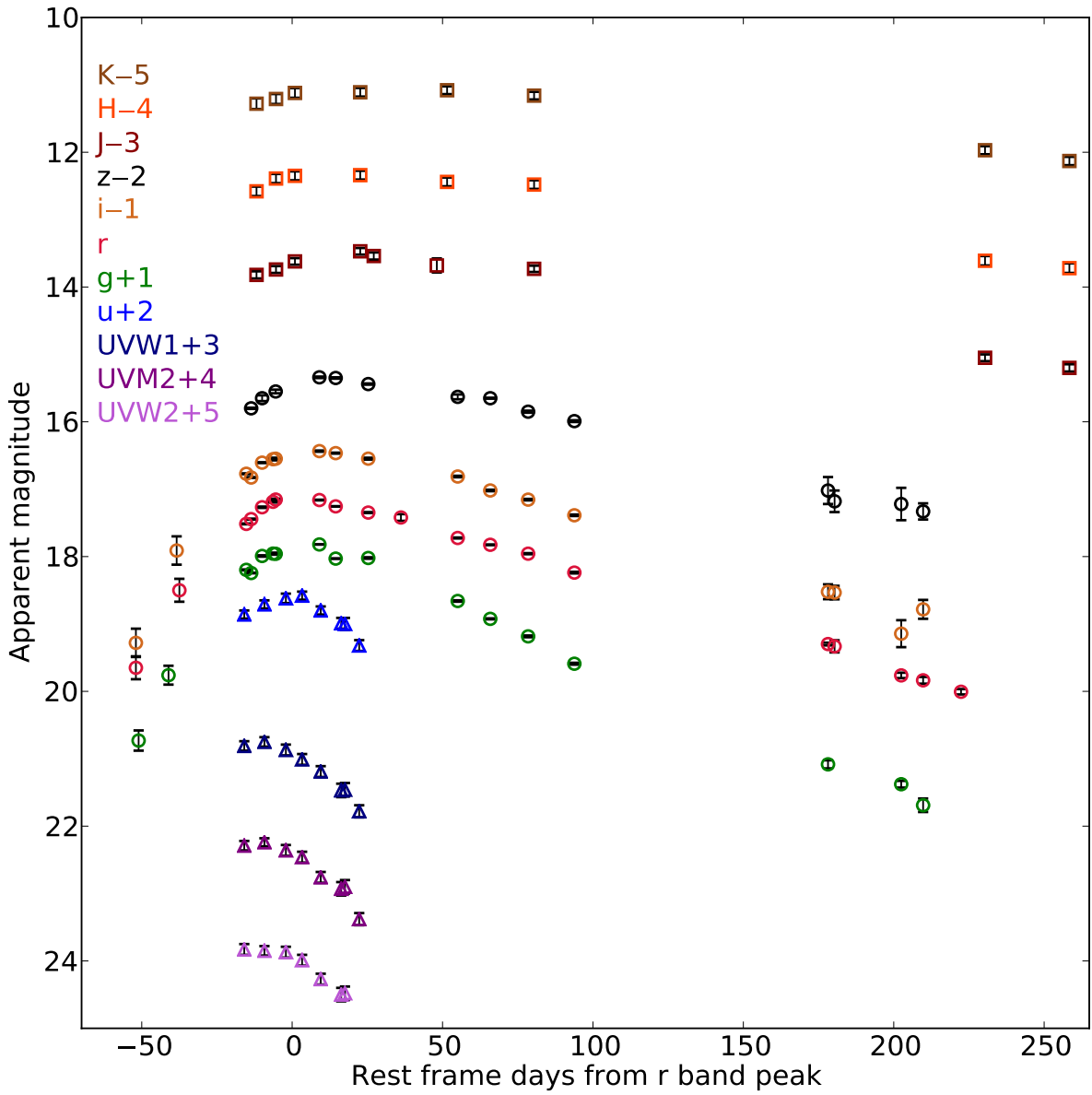

**Extended Data Fig. 1 | Multi-colour photometry of PTF12dam.** Observed light curve of PTF12dam in UVW2, UVM2, UVW1, *u*, *g*, *r*, *i*, *z* (AB magnitudes) and *J*, *H*, *K* (Vega magnitude system).

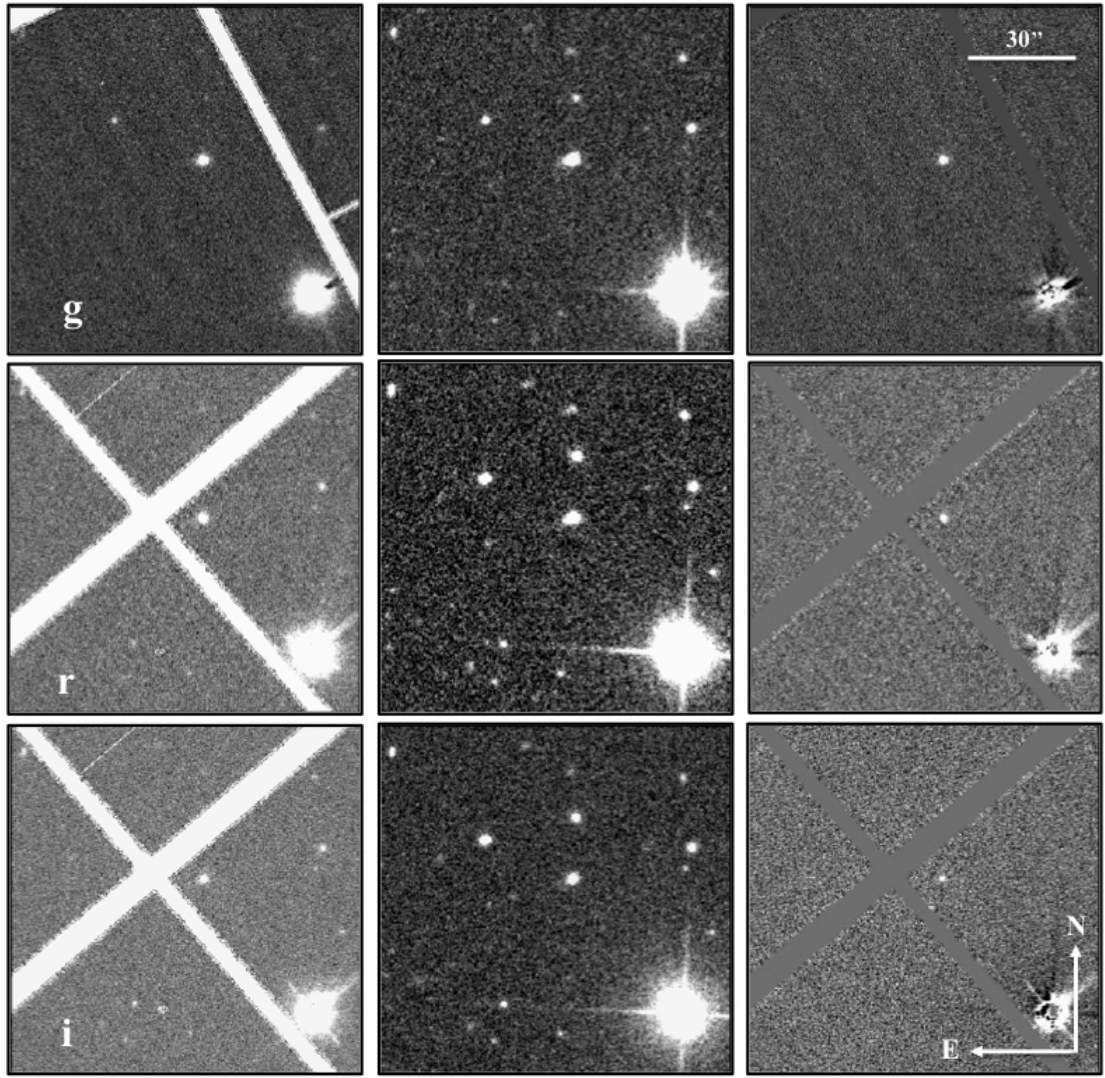

**Extended Data Fig. 2 | Image subtraction for the three earliest Pan-STARRS1 epochs of PTF12dam in $g_{P1}$, $r_{P1}$ and $i_{P1}$, using SDSS frames as reference images (taken on 11 February 2003).** These illustrate reliable image subtraction, resulting in clear detections of PTF12dam at early phases. The images on the left are our PS1 detections, those in the centre are the SDSS templates, and on the right are the differences between the two. The bright star in the lower right was saturated and hence does not subtract cleanly. At each PS1 epoch there are two images, taken as TTI (Transient Time Interval) pairs. Photometry was carried out and determined in the SDSS photometric system to match the bulk of the follow-up *griz* imaging. The white areas are gaps between the 590 × 598 pixel cells in the PS1 chip arrays.

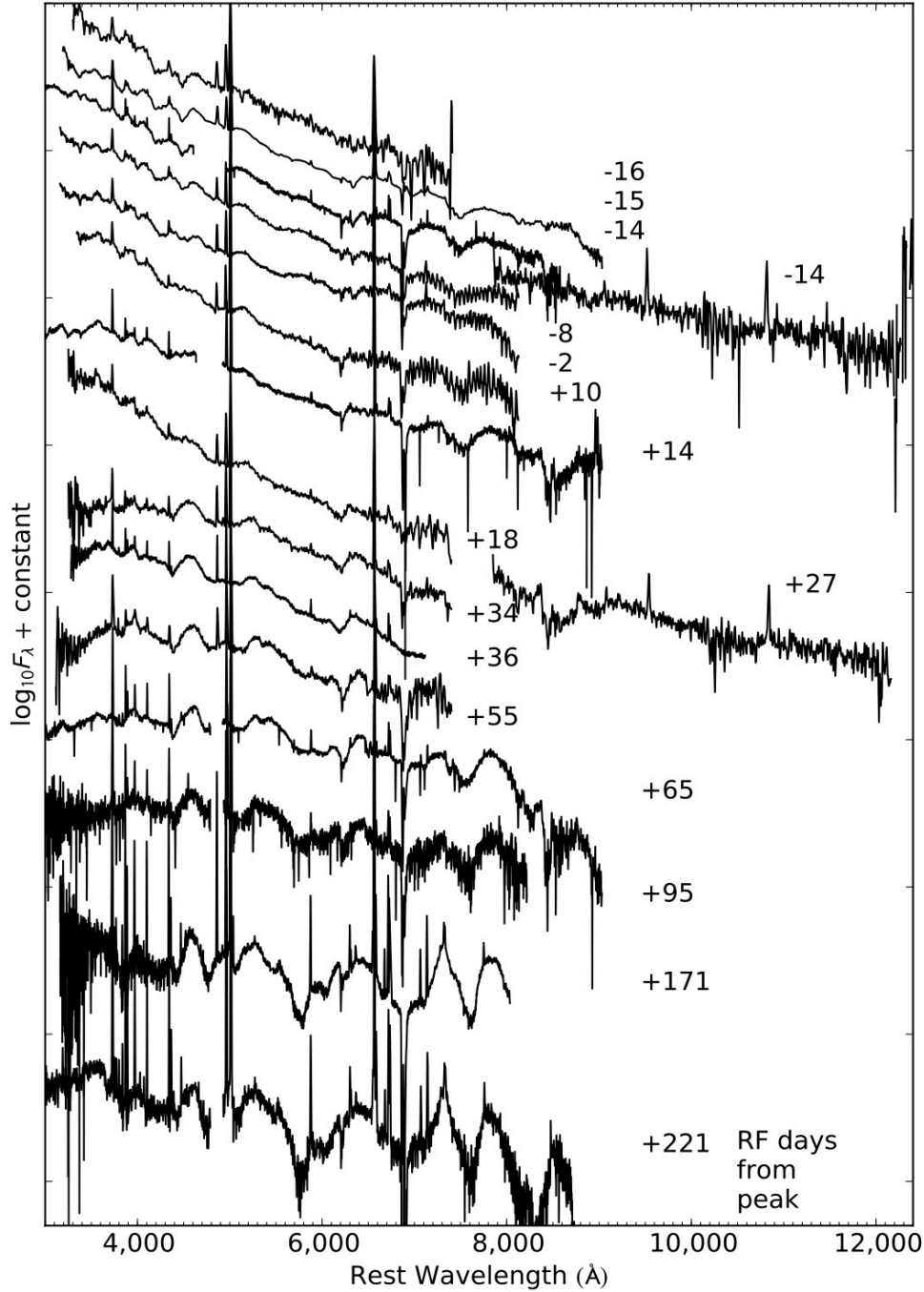

**Extended Data Fig. 3 | Spectral evolution of PTF12dam.** Full time-series optical and near-infrared spectroscopy of PTF12dam, from two weeks before maximum light to an extended pseudo-nebular phase at 100 to >200 days afterwards. A Starburst99 model continuum SED (spectral energy distribution) for the host galaxy has been calibrated against SDSS and GALEX (Galaxy Evolution Explorer) photometry, and subtracted from the last three spectra. RF, rest-frame.

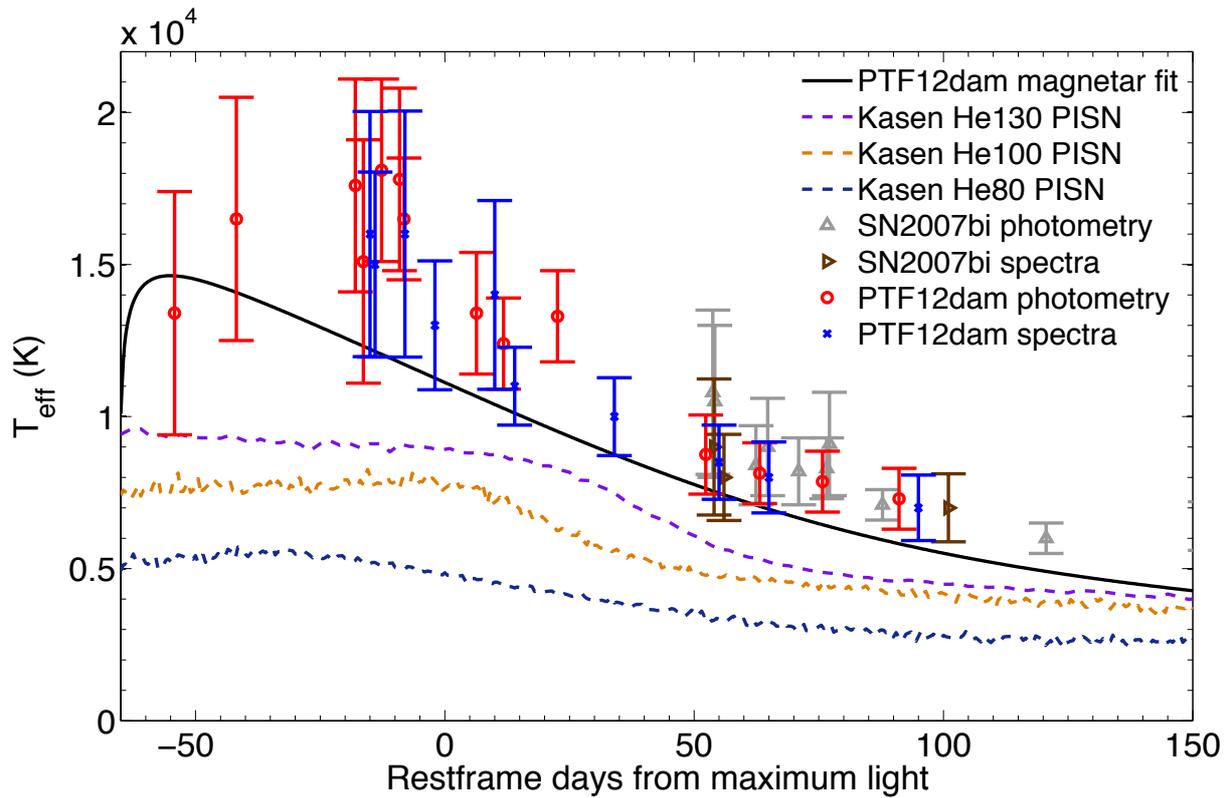

**Extended Data Fig. 4 | Effective temperature evolution of PTF12dam and SN 2007bi, compared with magnetar-powered and pair-instability models.** The magnetar model comes much closer to reproducing the high photospheric temperatures we observe, and matches the gradient of the decline phase well. PISN models do not reach such high effective temperatures, and show an approximately 100-day temperature plateau as they rise, before declining after maximum light.

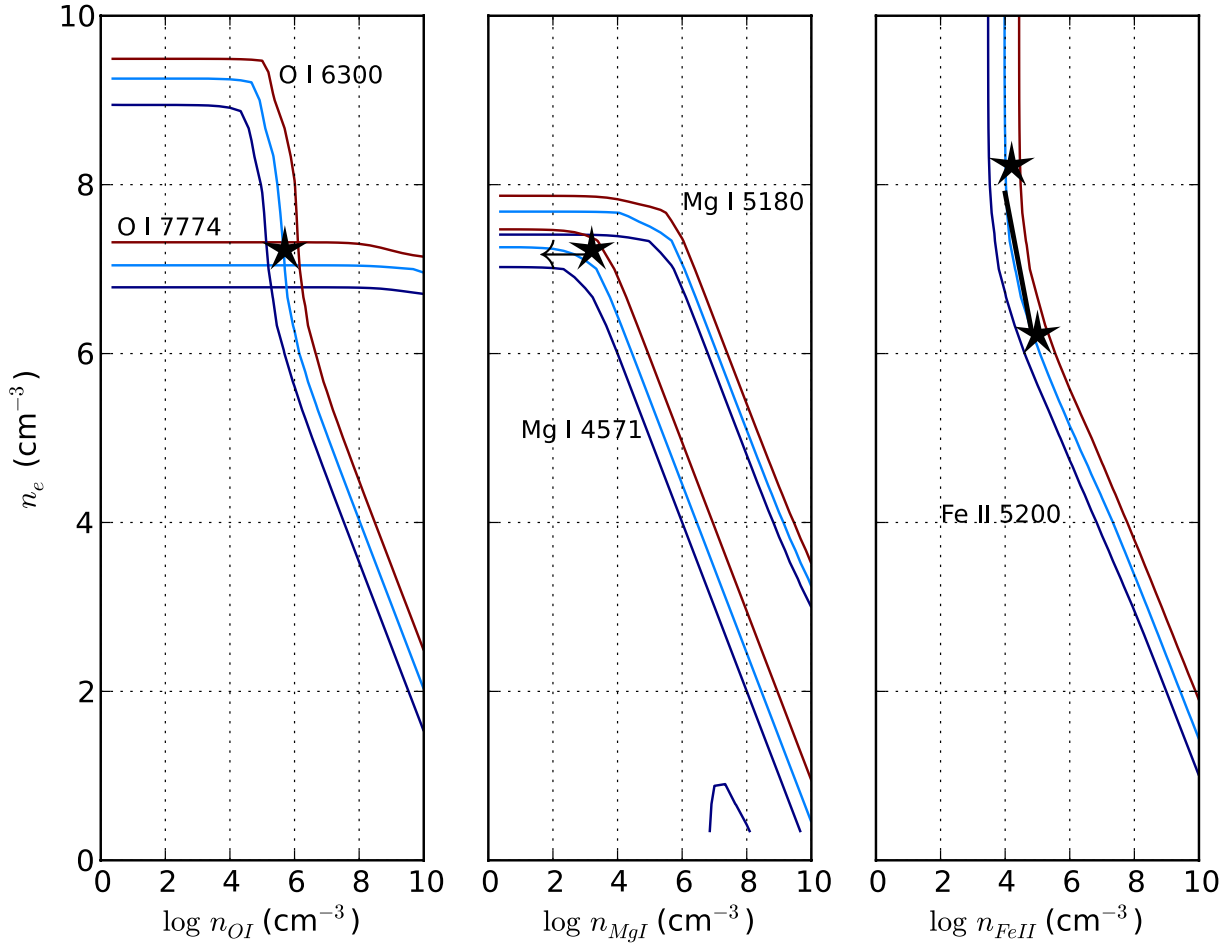

**Extended Data Fig. 5 | Modelling of the O I, Mg I and Fe II line fluxes in SN 2007bi at 367 days post-peak.** We plot contours for oxygen, magnesium and iron line fluxes predicted by our model in units of $L = 10^{40}$ erg s$^{-1}$ (dark blue = $L/3$; light blue = $L$; red = $3L$; where $L$ is the approximate luminosity of the lines in the 367-day post-peak spectrum of SN 2007bi) as functions of the respective ion density, $\{n_{\text{O I}}, n_{\text{Mg I}}, n_{\text{Fe II}}\}$, and electron density, $n_e$, at 5,000 K (approximately the temperature derived for the iron zone from the relative strengths of iron lines). The panels for O I and Mg I show two lines (O I 6,300, 7,774 Å; Mg I 4,571, 5,180 Å), whereas Fe II shows only contours for the 5,200 Å blend. No blending is likely to occur for any of the oxygen lines; the region where they intersect therefore gives the allowed densities, constraining $n_e$ to about $10^7$ cm$^{-3}$ (this is quite insensitive to the temperature we assume). Blending is also unlikely for Mg I] 4,571 Å, and the allowed Mg I density is therefore the intersection of this contour with $n_e \approx 10^7$ cm$^{-3}$, which can be seen to give $n_{\text{Mg I}} \lesssim 10^3$ cm$^{-3}$. At this magnesium density, we see that the Mg I 5,180 Å line makes some contribution to the 5,200 Å flux. Also shown is the allowed Fe II density at this temperature, for iron-zone electron densities spanning a factor of ten either side of that in the oxygen/magnesium zones.

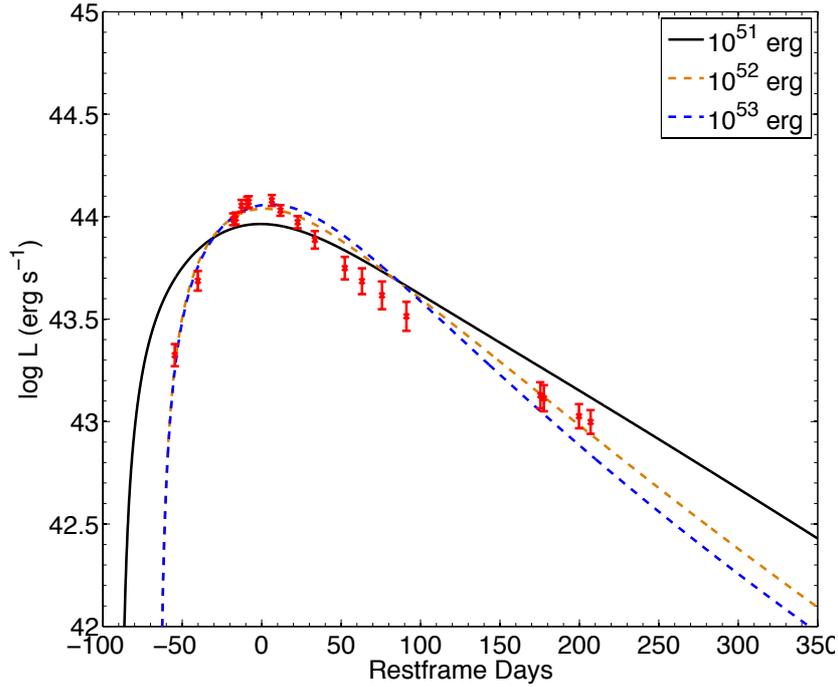

| Energy (erg) | Mass of $^{56}$Ni ($M_\odot$) | Ejecta Mass ($M_\odot$) | Total rise time (days) | $\chi^2$ / D.O.F. |
|---|---|---|---|---|
| $10^{51}$ | 14.1 | 14.7 | 88.9 | 8.7 |
| $10^{52}$ | 14.1 | 21.5 | 64.4 | 1.8 |
| $10^{53}$ | 16.1 | 52.2 | 64.4 | 3.0 |

**Extended Data Fig. 6 | Fits to the observed bolometric light curve of PTF12dam with radioactive $^{56}$Ni powered ejecta.** The formal fits of the models with kinetic energies of $10^{52}$ and $10^{53}$ erg are good (see graph), but the required combinations of $^{56}$Ni masses and ejecta masses (see data table) are not produced in physical models; such large nickel fractions are only expected to be produced in thermonuclear explosions (supernova Ia or possibly PISN), whereas the total ejected mass corresponds to the core-collapse of a massive star below the pair-instability threshold.

Page 22 of 26

| Date | MJD | RF Phase (days) | Telescope | g | $K_g$ | r | $K_r$ | i | $K_i$ | z | $K_z$ |
|---|---|---|---|---|---|---|---|---|---|---|---|
| 2012-04-13 | 56030.48 | -51.9 | PS1 | | | 19.45 (0.17)$^S$ | 0.20 | 20.04 (0.21)$^S$ | 0.24 | | |
| 2012-04-14 | 56031.45 | -51.0 | PS1 | 19.62 (0.15)$^S$ | 0.11 | | | | | | |
| 2012-04-25 | 56042.47 | -41.1 | PS1 | 18.65 (0.14)$^S$ | 0.11 | | | | | | |
| 2012-04-28 | 56045.48 | -38.4 | PS1 | | | | | 18.67 (0.21)$^S$ | 0.24 | | |
| 2012-04-29 | 56046.50 | -37.5 | PS1 | | | 18.30 (0.17)$^S$ | 0.20 | | | | |
| 2012-05-23 | 56071.13 | -15.2 | GTC + OSIRIS | 17.09 (0.01) | 0.11 | 17.32 (0.01) | 0.20 | 17.53 (0.01) | 0.24 | | |
| 2012-05-25 | 56072.92 | -13.6 | WHT + ACAM | 17.13 (0.01) | 0.11 | 17.26 (0.01) | 0.19 | 17.59 (0.01) | 0.24 | 17.52 (0.01) | 0.14 |
| 2012-05-29 | 56076.95 | -10.0 | LT + RATCam | 16.88 (0.01) | 0.11 | 17.11 (0.01) | 0.16 | 17.38 (0.01) | 0.23 | 17.37 (0.04) | 0.14 |
| 2012-06-02 | 56080.94 | -6.4 | LT + RATCam | 16.84 (0.01) | 0.12 | 17.05 (0.01) | 0.14 | 17.34 (0.02) | 0.22 | | |
| 2012-06-03 | 56081.94 | -5.5 | LT + RATCam | 16.84 (0.01) | 0.12 | 17.02 (0.01) | 0.13 | 17.33 (0.01) | 0.22 | 17.27 (0.03) | 0.14 |
| 2012-06-20 | 56098.03 | 9.1 | TNG + LRS | 16.76 (0.01) | 0.06 | 16.97 (0.01) | 0.19 | 17.24 (0.01) | 0.20 | 17.12 (0.01) | 0.11 |
| 2012-06-25 | 56104.02 | 14.5 | WHT + ACAM | 17.00 (0.01) | 0.03 | 17.05 (0.01) | 0.21 | 17.28 (0.01) | 0.19 | 17.15 (0.01) | 0.10 |
| 2012-07-07 | 56116.01 | 25.3 | NOT+ALFOSC | 17.05 (0.01) | -0.03 | 17.15 (0.01) | 0.19 | 17.37 (0.01) | 0.18 | 17.24 (0.01) | 0.10 |
| 2012-07-17 | 56128.03 | 36.2 | GTC + OSIRIS | | | 17.25 (0.05) | | | 0.17 | | |
| 2012-08-09 | 56148.93 | 55.0 | NOT+ALFOSC | 17.76 (0.01) | -0.10 | 17.51 (0.01) | 0.21 | 17.66 (0.01) | 0.15 | 17.51 (0.04) | 0.07 |
| 2012-08-21 | 56160.92 | 65.9 | WHT + ACAM | 17.99 (0.01) | -0.07 | 17.67 (0.01) | 0.16 | 17.88 (0.01) | 0.14 | 17.53 (0.01) | 0.05 |
| 2012-09-04 | 56174.85 | 78.5 | NOT+ALFOSC | 18.22 (0.01) | -0.03 | 17.86 (0.01) | 0.10 | 17.99 (0.01) | 0.16 | 17.75 (0.02) | 0.05 |
| 2012-09-21 | 56191.86 | 93.8 | WHT + ACAM | 18.58 (0.01) | 0.01 | 18.21 (0.01) | 0.03 | 18.20 (0.01) | 0.19 | 17.95 (0.01) | 0.02 |
| 2012-12-23 | 56285.21 | 178.1 | LT + RATCam | 19.84 (0.06)$^S$ | 0.24 | 19.27 (0.02)$^S$ | 0.03 | 19.47 (0.11)$^S$ | 0.05 | 19.03 (0.20)$^S$ | 0.01 |
| 2012-12-25 | 56287.63 | 180.3 | FTN + FS02 | | | 19.30 (0.09)$^S$ | 0.03 | 19.49 (0.01)$^S$ | 0.05 | 19.12 (0.11)$^S$ | 0.03 |
| 2013-01-19 | 56312.15 | 202.5 | LT + RATCam | 20.14 (0.05)$^S$ | 0.24 | 19.72 (0.04)$^S$ | 0.04 | 20.14 (0.20)$^S$ | 0.01 | 19.23 (0.24)$^S$ | -0.03 |
| 2013-01-27 | 56320.21 | 209.8 | LT + RATCam | 20.45 (0.10)$^S$ | 0.24 | 19.79 (0.05)$^S$ | 0.04 | 19.78 (0.14)$^S$ | 0.00 | 19.35 (0.12)$^S$ | -0.02 |
| 2013-02-10 | 56334.17 | 222.4 | LT + RATCam | | | 19.96 (0.04)$^S$ | 0.04 | | | | |
| 2003-02-11 | 52681.46 | | SDSS (host) | 19.30 (0.01) | | 19.15 (0.01) | | 18.70 (0.01) | | 19.31 (0.07) | |

**Extended Data Table 1 | Optical photometry of PTF12dam in SDSS *griz* bands, and *k*-corrections derived from our spectra.**
Magnitudes have been corrected for host galaxy contamination; those labelled with a superscript 'S' were determined after image subtraction with SDSS templates (see Supplementary Information section 2.1).

| Date | MJD | Phase | Telescope | UVW2 | UVM2 | UVW1 | SDSS u * | J | H | K |
|---|---|---|---|---|---|---|---|---|---|---|
| 2012-05-22 | 56070.38 | -15.9 | Swift+UVOT | 18.83 (0.08) | 18.29 (0.07) | 17.81 (0.07) | 16.86 (0.06) | | | |
| 2012-05-26 | 56074.89 | -11.8 | TNG+NICS | | | | | 16.82 (0.05) | 16.58 (0.07) | 16.28 (0.08) |
| 2012-05-30 | 56077.80 | -9.2 | Swift+UVOT | 18.85 (0.07) | 18.24 (0.06) | 17.75 (0.07) | 16.71 (0.06) | | | |
| 2012-06-03 | 56082.06 | -5.4 | TNG+NICS | | | | | 16.74 (0.05) | 16.39 (0.07) | 16.21 (0.07) |
| 2012-06-07 | 56085.69 | -2.1 | Swift+UVOT | 18.87 (0.08) | 18.36 (0.08) | 17.87 (0.08) | 16.62 (0.07) | | | |
| 2012-06-10 | 56089.03 | 0.9 | NOT+NOTCam | | | | | 16.62 (0.05) | 16.35 (0.07) | 16.12 (0.06) |
| 2012-06-13 | 56091.67 | 3.3 | Swift+UVOT | 18.99 (0.08) | 18.46 (0.08) | 18.01 (0.08) | 16.58 (0.06) | | | |
| 2012-06-20 | 56098.52 | 9.5 | Swift+UVOT | 19.27 (0.08) | 18.76 (0.08) | 18.19 (0.08) | 16.80 (0.06) | | | |
| 2012-06-27 | 56106.06 | 16.3 | Swift+UVOT | 19.50 (0.10) | 18.93 (0.10) | 18.47 (0.10) | 16.99 (0.08) | | | |
| 2012-06-28 | 56107.44 | 17.6 | Swift+UVOT | 19.48 (0.10) | 18.90 (0.10) | 18.46 (0.10) | 17.00 (0.09) | | | |
| 2012-07-04 | 56112.68 | 22.3 | Swift+UVOT | 20.16 (0.09) | 19.38 (0.09) | 18.78 (0.09) | 17.32 (0.08) | | | |
| 2012-07-04 | 56113.05 | 22.6 | NOT+NOTCam | | | | | 16.47 (0.05) | 16.34 (0.07) | 16.11 (0.06) |
| 2012-07-09 | 56118.04 | 27.1 | TNG+NICS | | | | | 16.54 (0.05) | | |
| 2012-08-01 | 56141.25 | 48.1 | UKIRT+WFCAM | | | | | 16.68 (0.10) | | |
| 2012-08-05 | 56145.03 | 51.5 | NOT+NOTCam | | | | | | 16.44 (0.07) | 16.08 (0.06) |
| 2012-09-07 | 56177.04 | 80.4 | NOT+NOTCam | | | | | 16.73 (0.05) | 16.48 (0.06) | 16.16 (0.06) |
| 2013-02-20 | 56343.00 | 230.4 | NOT+NOTCam | | | | | 18.05 (0.05) | 17.61 (0.07) | 16.97 (0.06) |
| 2013-03-22 | 56374.04 | 258.4 | NOT+NOTCam | | | | | 18.20 (0.06) | 17.72 (0.07) | 17.13 (0.06) |
| 2013-04-25 | 56407.00 | 288.2 | NOT+NOTCam | | | | | 18.22 (0.07) | 17.94 (0.08) | 17.25 (0.07) |

**Extended Data Table 2 | Photometry of PTF12dam outside the optical range.**
Ultraviolet photometry in Swift UVOT (Ultraviolet and Optical Telescope) bands, and near-infrared photometry in JHK (for details of the data, see Supplementary Information section 2.1).
*SDSS DR9 host magnitude: $u$ = 19.67 (0.03).

| Date | MJD | Telescope | $i_{P1}$ | $z_{P1}$ |
|---|---|---|---|---|
| 2010-12-31 | 55561.6 | PS1 | 21.49 (0.02) | |
| 2011-01-09 | 55570.55 | PS1 | 21.06 (0.02) | |
| 2011-01-15 | 55576.62 | PS1 | 20.74 (0.02) | |
| 2011-01-22 | 55583.52 | PS1 | | 20.78 (0.04) |
| 2011-01-24 | 55585.44 | PS1 | 20.47 (0.01) | |
| 2011-01-25 | 55586.61 | PS1 | | 20.70 (0.03) |
| 2011-01-28 | 55589.56 | PS1 | | 20.63 (0.03) |
| 2011-01-31 | 55592.58 | PS1 | | 20.58 (0.04) |
| 2011-02-03 | 55595.53 | PS1 | | 20.54 (0.02) |
| 2011-02-21 | 55613.44 | PS1 | | 20.38 (0.02) |
| 2011-03-11 | 55631.38 | PS1 | | 20.35 (0.03) |
| 2011-03-14 | 55634.32 | PS1 | | 20.41 (0.04) |
| 2011-03-26 | 55646.43 | PS1 | | 20.52 (0.03) |
| 2011-03-29 | 55649.50 | PS1 | | 20.56 (0.06) |
| 2011-04-22 | 55673.34 | PS1 | | 20.73 (0.03) |
| 2011-04-25 | 55676.34 | PS1 | | 20.81 (0.06) |
| 2011-05-01 | 55682.25 | PS1 | | 20.82 (0.04) |
| 2011-05-13 | 55694.28 | PS1 | | 20.90 (0.07) |
| 2011-05-22 | 55703.28 | PS1 | | 20.83 (0.06) |
| 2011-05-25 | 55706.26 | PS1 | | 20.99 (0.07) |
| 2011-05-31 | 55712.26 | PS1 | | 21.02 (0.03) |
| 2011-06-06 | 55718.26 | PS1 | | 21.08 (0.05) |
| 2011-12-30 | 55925.90 | PS1 | | 22.60 (0.14) |

**Extended Data Table 3 | Pan-STARRS1 photometry of PS1-11ap used in this work.**
The $i_{P1}$ magnitudes are transformed to $z_{P1}$ using the observed colour $i - z = -0.18$ at the earliest $z$ point, MJD = 55583.52, with $i$ linearly interpolated to this epoch (see Supplementary Information section 2.2).

| Date | MJD | RF phase (days) | Instrument | Grism/Grating | Range (Å) | Resolution (Å) |
|---|---|---|---|---|---|---|
| | | | PS1-11ap | | | |
| 2011-02-22 | 55614 | -1 | WHT + ISIS | R300B; R158R | 3150-10500 | 12 |
| 2011-06-22 | 55734 | +78 | GN + GMOS | R150 | 4000-11000 | 23 |
| | | | PTF12dam | | | |
| 2012-05-23 | 56070.99 | -16 | Asiago Copernico + AFOSC | Gr04 | 3400-8200 | 25 |
| 2012-05-24 | 56071.12 | -15 | GTC + OSIRIS | R300R | 3500-10000 | 30 |
| 2012-05-25 | 56072.91 | -14 | WHT + ISIS | R300B; R158R | 3250-5100; 5500-9500 | 3;5 |
| 2012-05-26 | 56073.95 | -13 | TNG + NICS | IJ | 8700-14500 | 35 |
| 2012-06-01 | 56079.96 | -8 | NOT + ALFOSC | Gr04 | 3500-9000 | 15 |
| 2012-06-08 | 56086.95 | -2 | NOT + ALFOSC | Gr04 | 3500-9000 | 15 |
| 2012-06-21 | 56100.04 | +10 | NOT + ALFOSC | Gr04 | 3700-9000 | 15 |
| 2012-06-25 | 56103.99 | +14 | WHT + ISIS | R300B; R158R | 3200-5200; 5450-10000 | 6;11 |
| 2012-06-29 | 56107.97 | +18 | Asiago Copernico + AFOSC | VPH6 | 3600-10000 | 15 |
| 2012-07-09 | 56117.99 | +27 | TNG + NICS | IJ | 8700-13500 | 35 |
| 2012-07-17 | 56125.95 | +34 | Asiago Copernico + AFOSC | Gr04 | 3900-8140 | 13 |
| 2012-07-19 | 56128.03 | +36 | GTC + OSIRIS | R1000B | 3600-7900 | 7 |
| 2012-08-09 | 56148.95 | +55 | NOT + ALFOSC | Gr04 | 3500-8200 | 20 |
| 2012-08-20 | 56159.92 | +65 | WHT + ISIS | R300B; R158R | 3200-5300; 5450-10000 | 4;7 |
| 2012-09-22 | 56192.86 | +95 | WHT + ISIS | R300B; R158R | 3200-5300; 5450-10000 | 6;11 |
| 2012-12-16 | 56277.5 | +171 | GN + GMOS | B600; R400 | 3500-8900 | 3;4 |
| 2013-02-10 | 56334.17 | +221 | WHT + ISIS | R300B; R158R | 3200-5300; 5400-10000 | 5;10 |

**Extended Data Table 4 | Log of spectra for PTF12dam and the PS1-11ap spectra used in this work.**
Spectral evolution of PTF12dam is plotted in Extended Data Fig. 3.. Full PS1-11ap time-series are to be presented[31].

# Corrigendum: Slowly fading super-luminous supernovae that are not pair-instability explosions

We have identified an important error in Nicholl et al. (2013), Nature 502, 346. This error affects Figure 4 and Extended Data Figure 6, as well as the values of some parameters derived from our model fits. We stress that this error in no way affects any of the discussion presented in the paper or the conclusions drawn.

The error is as follows. In building the bolometric light curve of the superluminous supernova PTF12dam, our code assumed that photometry from the *Swift* satellite was calibrated in the Vega magnitude system. However, our photometry was actually calibrated to the AB magnitude system (and published in the AB system in our original paper). This led to an underestimate of ~50% in the measured peak luminosity of PTF12dam.

Here we present updated figures and model fits with the correct bolometric luminosity. To construct the bolometric light curve, we transformed the *Swift* data into Vega magnitudes, and then converted all photometry to fluxes. At epochs with the full range of *UVW2* to *K* band, we simply integrated over the observed SED. At epochs with missing filters, we accounted for the unobserved flux by fitting blackbodies to the available data. We also compared our blackbody extrapolations against polynomial fits to the UV and NIR light curves, finding consistent results. This should be more reliable than our previous extrapolation method, which assumed linear colour evolution over 40 days.

If anything, the improved bolometric light curve strengthens our main conclusion — that PTF12dam was not a pair-instability supernova — as the brighter light curve peak results in an even steeper rise to maximum. It is important to note that the large discrepancy compared to pair-instability models does not rely solely on bolometric comparisons: the difference was clearly apparent in the *r*-band light curves in our original Figure 1. Thus this is a robust result independent of any time-varying bolometric correction.

Our secondary conclusion — that spin down of a nascent magnetar can satisfactorily explain the observed properties — also remains intact. The parameters of our magnetar-powered fit to the corrected bolometric light curve shown in Figure 4 are: magnetic field $B = 5\times10^{13}$ G; spin period $P = 2.3$ ms; and ejecta mass $M_{ej} = 7$ M$_\odot$ for an opacity $\kappa = 0.1$ cm$^2$ g$^{-1}$ and explosion energy $E = 10^{51}$ erg. Thus these parameters remain within a sensible range. Our suggestion that a relatively lower spin period and larger ejected mass can explain the existence of these long-duration superluminous supernovae is unchanged.

In Extended Data Figure 6, we showed that decay of radioactive nickel-56 could not explain the observed light curve. This remains true for the corrected light curve. The unrealistic parameters required to model the data with nickel as the power source are listed below the figure.

We thank P. Vreeswijk for initially pointing out a discrepancy between our light curve and his own results. M. Nicholl identified the source of the discrepancy.

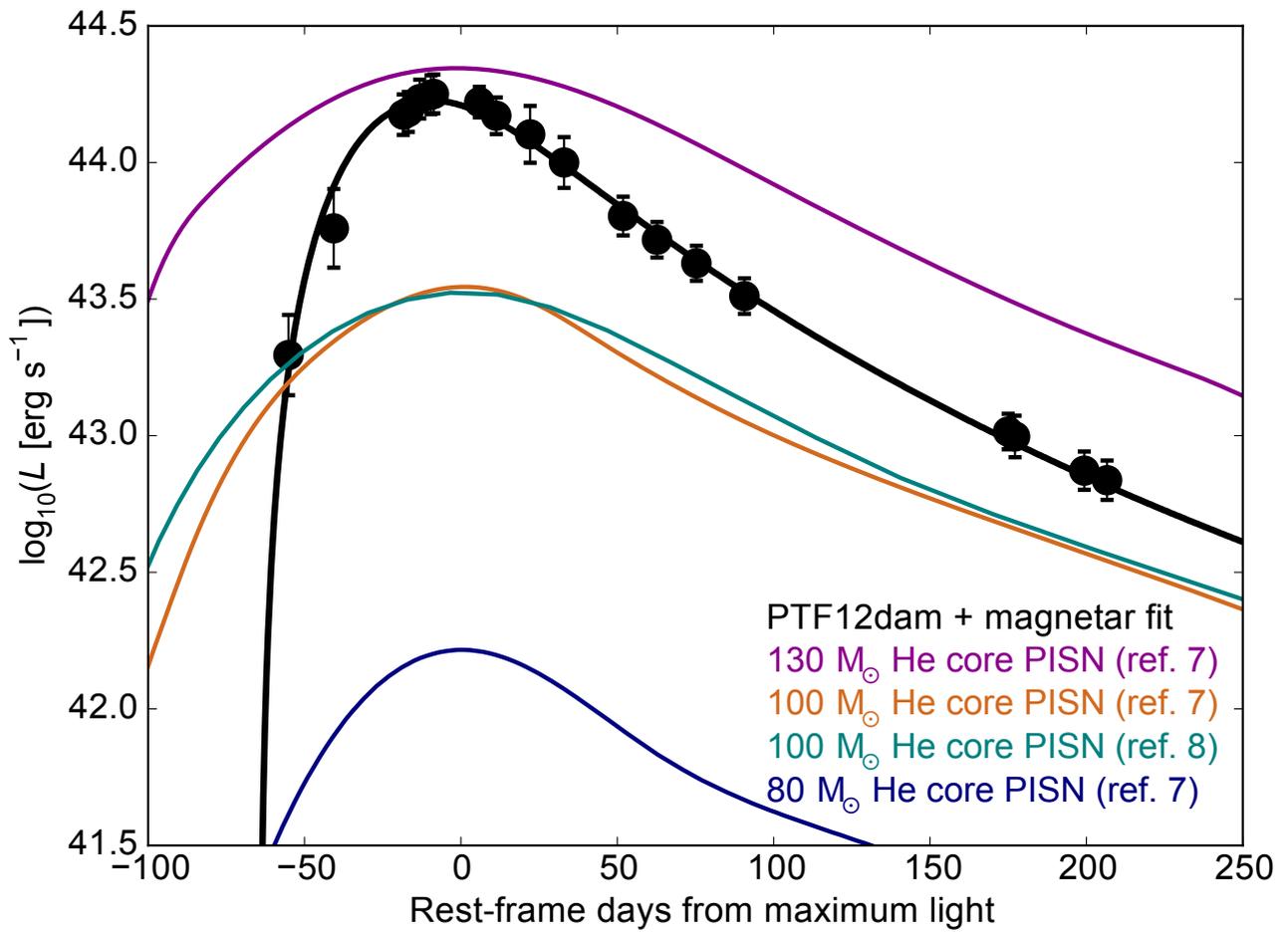

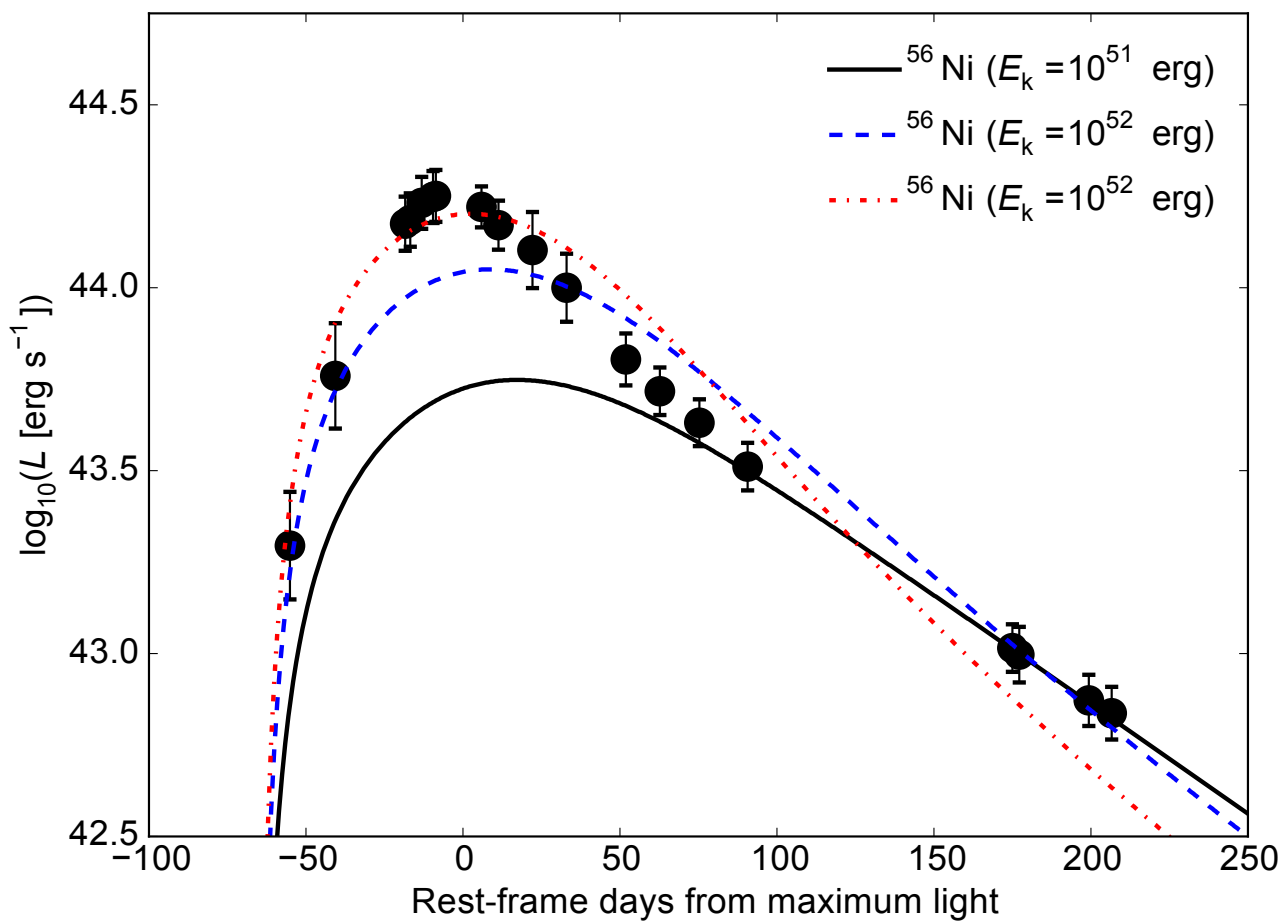

| Energy (erg) | Mass of $^{56}$Ni (M$_\odot$) | Ejecta mass (M$_\odot$) | $\chi^2$/D.O.F. |
|---|---|---|---|
| $10^{51}$ | 8.5 | 8.6 | 23.8 |
| $10^{52}$ | 16.7 | 16.8 | 4.1 |
| $10^{53}$ | 25.0 | 35.5 | 2.5 |

# Supplementary Information

## 1. Imaging observations and data reduction

### 1.1 Pan-STARRS1 imaging for PTF12dam (= PS1-12arh) and PS1-11ap

The Pan-STARRS1 (PS1) system and its application to transients has been described in refs 21, 23, 24, 25, 32 and 33. We summarize the relevant system details here for completeness.

PS1 is a high-etendue wide-field imaging system designed for dedicated survey observations, on a 1.8 meter telescope on Haleakala with a 1.4 Gigapixel camera and a 7 deg$^2$ field of view. The PS1 observations are obtained through a set of five broadband filters, which are $g_{P1}$ ($\lambda_{eff}$ = 483 nm), $r_{P1}$ ($\lambda_{eff}$ = 619 nm), $i_{P1}$ ($\lambda_{eff}$ = 752 nm), $z_{P1}$ ($\lambda_{eff}$ = 866 nm), and $y_{P1}$ ($\lambda_{eff}$ = 971 nm). See ref. 33 for full details of the bandpasses.

This paper uses images and photometry from both the PS1 Medium Deep Field survey (MDS) and the wide 3π Survey. The goal of the 3π Survey is to observe the portion of the sky North of -30 deg declination, with a total of 20 exposures per year across all five filters for each field center. The 3π survey plan is to observe each position 4 times in each of $g_{P1}$, $r_{P1}$, $i_{P1}$, $z_{P1}$, and $y_{P1}$ during a 12 month period, although this can be interrupted by weather. The 4 epochs in a calendar year are typically split into two pairs called Transient Time Interval (TTI) pairs, which are single observations separated by 20-30 minutes to allow for the discovery of moving objects. The exposure times at each epoch (i.e. in each of the TTI exposures) are 43s, 40s, 45s, 30s and 30s in $g_{P1}$, $r_{P1}$, $i_{P1}$, $z_{P1}$ and $y_{P1}$, leading to 5σ depths of roughly 22.0, 21.6, 21.7, 21.4. and 19.3 (in the PS1 AB system described by ref. 32). The PS1 images are processed by the Pan-STARRS1 Image Processing Pipeline (IPP), on a computer cluster hosted in the Maui High Performance Computer Center. This performs automatic bias subtraction, flat fielding, a flux-conserving warping to a sky-based image plane, masking and artifact removal, object detection, photometry and astrometry[34,35]. The TTI pairs are not stacked together, but kept as individual frames. Full stacking of all data across the sky, over the three years is now underway but for the purposes of transient searches, the individual exposures are kept separate.

The PS1 MDS obtains deep multi-epoch images in the $g_{P1}$, $r_{P1}$, $i_{P1}$, $z_{P1}$ and $y_{P1}$ bands of 10 fields with a typical cycle of observations being $g_{P1}$ and $r_{P1}$ on one night, followed by $i_{P1}$ and $z_{P1}$ bands on the subsequent nights[23]. In some cases this cycle is broken to optimise for sky brightness. Observations in the $y_{P1}$ band are taken close to the full moon. The MDS Images are also processed through the Image Processing Pipeline, and are stacked to give a single nightly image containing eight exposures in a dithered sequence. The observing season for each field is 6 months per year. PS1-11ap was detected throughout the 2011 observing season for MD05.

**1.2 Follow-up imaging for PTF12dam**

Optical imaging in SDSS-like g, r, i and z filters was obtained with RATCam on the 2.0m Liverpool Telescope, and FS02 on the 2.0m Faulkes Telescope North. The data were automatically reduced by respective facility pipelines to produce detrended images (bias and flat field corrected). Optical images obtained using ACAM on the 4.2m William Herschel Telescope, OSIRIS on the 10.4m Gran Telescopio Canarias, LRS on the 3.58m Telescopio Nazionale Galileo, and ALFOSC on the 2.56m Nordic Optical Telescope (NOT) were reduced using standard tasks within the IRAF [i] package CCDRED, to debias, trim and flatfield the images. Multiple exposures from the same epochs and filters were median-combined with cosmic ray rejection using the IMALIGN and IMCOMBINE tasks.

Near infrared imaging was taken from three sources: NOTCam on the NOT, NICS on the TNG, and WFCAM on the UK infrared telescope. NOT and UKIRT data were reduced by facility pipelines, while we reduced TNG data using standard IRAF packages. Images were flat field corrected and sky subtracted. For each position in the mosaic making up the image, a sky frame was created by median combining exposures from all the other positions in the dithering pattern to get rid of stars. The sky-subtracted individual dithers were aligned and combined to produce the final images.

UV observations were obtained with UVOT on board the Swift satellite. The frames were reduced using tools within HEAsoft[ii]. UVOT data were obtained in uvw2, uvm2, uvw1 and u filters, with spatial resolution of about 2 arcsec (full width at half maximum, FWHM). Eight epochs, spread over a period of 50 days, are available. Individual images for each epoch were first co-added, before aperture magnitudes were measured following the prescription of ref. 36.

---

[i] Image Reduction and Analysis Facility (IRAF), a software system distributed by the National Optical Astronomy Observatories (NOAO).
[ii] Available from the NASA High Energy Astrophysics Science Archive Research Center

## 2. Photometric measurements

### 2.1 PTF12dam in PS1 data and follow-up data

The host galaxy of PTF12dam is bright ($g$ = 19.30, or $M_g$ = −19) compared to most ultra-luminous supernova hosts[37] and we correct for galaxy flux as follows. For observations in g,r,i,z where PTF12dam+host is more than a magnitude brighter than the host alone, we subtract the contribution from the host flux in a given band simply using the magnitudes (model mags) from SDSS DR9 (ref. 38). The host is a compact source, such that essentially all of the flux falls within the SN PSF. When the brightness of the supernova and galaxy are comparable, we subtract an SDSS template image from our target image using the HOTPanTS[iii] code (ref. 39), which matches the seeing in the two images by computing a convolution kernel from the PSFs of point sources in the images. The resultant subtracted image contains only flux from the supernova. Extended Data Fig. 2 shows the image subtractions for the critical PS1 early detections, which constrain the rise time.

In both cases, photometric flux measurements were performed using the custom built SNOOPY package[40] (implemented in IRAF by E. Cappellaro). This suite of programs is based on the standard DAOPHOT PSF-fitting task, available within IRAF. The zero point for each observation was calibrated by comparing multiple point sources in the field with SDSS photometry. Colour corrections and extinction for each site were then used to refine the measured magnitudes.

As PTF12dam is at a redshift of $z$ = 0.107, a $k$-correction was applied to convert our measured magnitudes in each filter to the magnitudes that would be obtained in the restframe. To compute the $k$-corrections, we used SYNPHOT[iv], to calculate synthetic $gri$ photometry at every epoch for which we obtained a spectrum. Magnitudes (in the SDSS AB system) were measured for the observed spectrum and after correcting it to restframe. The $k$-corrections (shown in 1) were thus the differences between these two, and were then applied to the measured photometry from the imaging. The corrections for other epochs were calculated using linear interpolation. The z band lies at the red edge of the optical spectra, and hence was not covered by both the observed and rest frame spectra simultaneously. For these magnitudes, colour-based $k$-corrections were used instead[v] (refs 41, 42).

---

[iii] http://www.astro.washington.edu/users/becker/hotpants.html
[iv] SYNPHOT is a product of the Space Telescope Science Institute, which is operated by AURA for NASA.
http://www.stsci.edu/institute/software_hardware/stsdas/synphot
[v] http://kcor.sai.msu.ru/

The errors were calculated in SNOOPY using an artificial star experiment. The fitted supernova PSF is placed at different locations on the image, and the magnitude is computed each time. The standard deviation of these measurements gives the error in fitting the background, which is the dominant source of error since the bright supernova has a well-defined PSF that can be fit at all times.

Table 1 lists the *griz* ground-based photometry after image subtraction (or host flux subtraction), and before the *k*-correction is applied. The absolute AB magnitudes in these filters are calculated using a flat ΛCDM cosmology with $H_0 = 72$ km s$^{-1}$ Mpc$^{-1}$, $\Omega_M = 0.27$ and $\Omega_\Lambda = 0.73$. The date of maximum light (MJD = 56088, corresponding to 10$^{th}$ June 2012) was determined from our bolometric light curve (Section 5); this coincides with the peak in *r*. Epochs are taken relative to this, and corrected for cosmological time dilation using the observed redshift $z = 0.107$.

Swift photometry was carried out using aperture photometry with the HEAsoft tools, and calibrated using the latest calibration database provided by HEASARC. A secondary calibration was carried out by calculating the mean shifts in brightness of close stars from their average magnitudes in each band. To transform the UVOT *u* magnitudes from the instrumental system into standard SDSS magnitudes (in order to maintain consistency with our optical imaging, and ease the creation of our bolometric light curve; see section 5), a shift of $\Delta u = -0.21$ mag was applied. This shift was computed from the magnitudes of stars in the PTF12dam field listed in SDSS. Thus, Table 2 lists the SWIFT photometric measurements in the AB system. Host galaxy flux, inferred from our model host template (see section 3) was subtracted from all UVOT photometry. Due to the lack of UV spectral coverage, no *k*-correction has been applied to the SWIFT magnitudes.

Near infrared photometry was carried out using aperture photometry with the PHOT task in IRAF, and calibrated using the magnitudes of nearby sources in the 2MASS catalogue. Because only a few 2MASS sources were available in the field, a secondary calibration was carried out. In each band, we calculated the mean magnitudes of sources with flux similar to that of the supernova. The difference from their mean values (averaged over these sources) in each image was measured; this shift was then applied to the calculated supernova magnitude. No image subtraction was applied for the NIR photometry. There is no host detected in 2MASS, to limiting magnitudes of $J = 17$, $H = 16$, $K = 15.5$. No *k*-correction was applied to the NIR data, as we lack sufficient NIR spectral coverage to reliably cover all epochs. Photometric measurements, in Vega magnitudes, are given in Table 2.

All PTF12dam photometry is plotted in Extended Data Figure 2.

## 2.2 PS1-11ap in the Pan-STARRS1 Medium Deep Survey

For the MDS, the 8 images taken during any one night are stacked to produce a "nightly stack". This nightly data product is used in two image differencing pipelines that run simultaneously, but independently. The PS1 system is developing the Transient Science Server (PS1 TSS), which automatically takes the nightly stacks, creates image differences with manually created deep reference images, carries out PSF fitting photometry on the image differences, and returns catalogues of variables and transient candidates. Photometric and astrometric measurements are performed by the PS1 IPP system at the Maui High Performance Computing Centre and ingested into a MySQL database hosted at Queen's University Belfast. Independent difference image analysis is also run with the *photpipe* pipeline[43] hosted at Harvard/CfA, and since this uses forced photometry and an accurate zeropoint calibration, we employ the *photpipe* measurements for PS1-11ap in this paper. This pipeline produces image differences from the IPP-created nightly stacks, with respect to a custom-built deep reference stack. Forced-centroid PSF-fitting photometry is applied on its image differences, with a PSF derived from reference stars in each nightly stack. The zeropoints were measured for the AB system from comparison with field stars in the SDSS catalog. The Poisson error is propagated through the resampling and image differencing. Since this does not take the covariance into account, *photpipe* also runs forced photometry in apertures at random positions and calculates the standard deviation of the ratio between the flux and the error. All errors are multiplied by the standard deviation to correct for the covariance. The difference imaging photometry in the observer frame $z_{P1}$ band is reported in AB magnitudes in Table 3. This was corrected to an absolute restframe AB mag at $\lambda_{\text{eff}} = 5680$ Å using

$$M_{5680} = z_{P1} - 5\log\left(\frac{d_L}{10}\right) + 2.5\log(1+z)$$

where $z_{P1}$ is the apparent AB magnitude in the $z_{P1}$ filter, $d_L$ is the luminosity distance in parsecs, and $z$ is the redshift. This equation corrects for cosmological expansion, but is not a full *k*-correction. $M_{5680}$ should be close to the absolute r-band AB magnitude. We have earlier coverage in the PS1 $i_{P1}$ band, so to compare the rise we convert $i_{P1}$ to $z_{P1}$ using the observed colour at the earliest $z_{P1}$ point, $i_{P1} - z_{P1} = -0.18$. The full dataset of $g_{P1}$, $r_{P1}$, $i_{P1}$, $z_{P1}$ and $y_{P1}$, with full *k*-corrections, supplementary data, and spectral series, will be presented in a companion paper[31]. These absolute *r* magnitudes are plotted in Figure 1.

## 2.3 SN2007bi photometry

The published photometry for SN2007bi[1,10] has until now been in the Johnson-Cousins *UBVRI* system. For ease of comparison, we converted the existing R band photometry to SDSS *r* magnitudes. This was done using colour transformations in *V–R* (ref. 44) between SDSS and Johnson-Cousins filter systems. Multicolour photometry of SN2007bi was used where available[10] although most photometric data for SN2007bi is in *R* only[1]. We considered linear interpolation of the required correction between these epochs, but found that the conversion factor varied negligibly with time, and so were able to apply a constant shift of +0.15 magnitudes to the *R* data, to bring it into the AB *r* system, and estimated the additional error due to the uncertainty in *V* magnitudes and the coefficients in the transformation. While a detailed S-correction would have been preferable, the lack of SN2007bi spectra earlier than 54d post-peak meant that such a correction would necessarily have been based on the spectra of PTF12dam. We felt that the additional uncertainty this would introduce (to what is only a small correction) meant that it was not worthwhile.

## 3 Spectroscopic data and analysis of PTF12dam and PS1-11ap

Optical spectra of PTF12dam were taken with the Gran Telescopio Canarias (GTC) with the OSIRIS instrument; the William Herschel Telescope plus ISIS spectrograph; the Nordic Optical Telescope plus ALFOSC; the Asiago Copernico Telescope plus AFOSC; and the Gemini North telescope with GMOS-N. Spectra of PS1-11ap were taken with the William Herschel Telescope plus ISIS and the Gemini North telescope with GMOS-N. The details of the wavelength coverage and resolution are listed in Table 4.

Standard procedures within IRAF were used to detrend the CCD data, and extraction of spectra was carried out using variance-weighted cleaning with the IRAF task APALL. When this was insufficient to remove cosmic rays, the 2D frames were first cleaned using LACosmic[vi] (ref. 45). Spectra were wavelength-calibrated using spectra of arc lamps for comparison, and flux-calibrated using sensitivity functions derived from the spectra of standard stars obtained on the same nights as our spectra.

Observed spectra were adjusted to restframe by applying a redshift correction, and were also corrected for extinction. Prominent host galaxy lines (in particular Hα, Hβ and the [O III] 4959 & 5007 Å doublet) gave $z = 0.107$. We measured the flux ratio of Hα/Hβ as 2.99 in the observed frame in the NOT spectrum from 2012 August 9$^{th}$. This compares to an expected intrinsic

---

[vi] http://www.astro.yale.edu/dokkum/lacosmic/

line ratio of 2.86 for case B recombination[46]. We assumed that $R_v$ for this host galaxy is similar to that observed in the LMC ($R_v$ = 3.16; ref. 47), and hence we estimated a host galaxy extinction in the V band of $A_V^{host}$ = 0.1 mag from the Balmer decrement. Within the uncertainties, we find similar results if we were to apply $R_v$ = 2.93, which has been proposed for an SMC-like environment[47]. We used a Milky Way extinction in the direction of PTF12dam of $A_V^{MW}$ = 0.037 mag from the NASA/IPAC IRSA dust maps[vii] (ref. 48). These were applied separately to the spectra and to all filters with the appropriate redshift corrections. Host reddening was assumed to be negligible for PS1-11ap and SN2007bi, and only a galactic extinction correction was applied.

Near infrared spectra were obtained using NICS on the TNG on 26th May and 9th July 2012. As for the optical, the spectra were calibrated in wavelength through spectra of comparison lamps acquired with the same configuration of the PTF12dam observation. First order flux calibrations were obtained using A0 standard stars taken in the same night with the same set-up used for PTF12dam. Solar analogues at a similar airmass were observed either before or after PTF12dam, to facilitate the removal of the strong telluric absorptions between 1 and 2 μm. The spectra show no broad hydrogen or helium (consistent with the optical spectra). The early (-13d) spectrum is nearly featureless with two narrow host galaxy lines from [S III] (9500 Å) and He I (10580 Å), while the later one, at 27 days after peak, shows broad absorption due to the Ca II NIR triplet, which is also seen in the optical spectra with the longest wavelength coverage.

The two latest optical spectra of PTF12dam, at 171d and 221d after peak, have significant contamination from the host galaxy in the continuum, as judged from the pre-discovery SDSS flux of the host and our photometric measurements. The host is not resolved from the SN hence no host subtraction is possible in the 2D spectroscopic reductions. At the present time, we do not have a spectrum of the host hence we constructed a galaxy template to subtract. We used *starburst99*[viii] (ref. 49) to calculate series of spectra for both continuous star-formation and an initial burst, settling on a 30 Myr old stellar population with a continuous star-formation history (at a metallicity of 0.05 solar, and a Salpeter initial mass function). This does not include nebular emission lines, hence we added narrow emission lines with fluxes as measured from the spectra of PTF12dam at 221d (after continuum subtraction). This provided a galaxy template spectrum, which we scaled and reddened until synthetic photometry (with the IRAF task SYNPHOT) through SDSS ugriz and GALEX FUV and NUV filters matched pre-discovery measurements[38]. Thus we have a model galaxy spectrum that reproduces the observed flux. The continuum from this model spectrum was then subtracted from the PTF12dam spectra and NUV photometry in Figs 2 and 3. With this subtraction, the nebular features are more prominent, as one would expect. In fact, at this phase, the +171d pseudo-nebular spectrum looks almost identical

---

[vii] http://irsa.ipac.caltech.edu/applications/DUST/
[viii] http://www.stsci.edu/science/starburst99/docs/default.htm

to SN2007bi after +134d. A similar fit was made for the host of PS1-11ap, and subtracted accordingly.

All spectra of PTF12dam are shown in Extended Data Fig. 3.

At the epochs after peak luminosity, while PTF12dam and PS-11ap are still in the photospheric phase, the spectral lines are assumed to be those identified for SN2007bi[1], as the spectra are closely matched. However the early spectra we have of these two super-luminous SNe before and around maximum light explore new epochs, particularly in the near-UV. We determine the main features of our spectra before maximum light using SYN++, a C++ version of the commonly used synthetic spectrum tool SYNOW[ix] (refs 50, 51). Single-ion spectra were generated for common ions in ejecta with temperatures and velocities appropriate to our spectra ($T \sim 15000$ K and $v \sim 11000$ km s$^{-1}$). We find that all of the main features can be accounted for with O II, Ca II, Fe III, Mg II and Si II.

## 4  Nebular phase modelling

Nebular phase modelling[1] of the SN2007bi spectrum was a key component in the argument for a large ejecta mass and PISN explanation. The model achieved a good fit to most of the lines, but crucially the strong [Fe II] 7155 Å line predicted in the model does not appear in the observed spectrum. The only strong iron line identified was an [Fe II] 5200 Å blend. It was the strength of this line that possibly suggested[1] a high mass of $^{56}$Ni, in combination with an ejecta mass of 60-80M$_\odot$, was required. While this is a self-consistent argument in favour of the PISN scenario, the aim of this section is to explore if it is the only plausible solution.

Using the NLTE solver[19], we investigated whether the ejecta parameters of our proposed core-collapse and magnetar-heating scenario are consistent with the nebular spectra of SN2007bi. PTF12dam appears to be evolving to a similar spectrum, although our last spectrum is still only quasi-nebular. Full spectral modelling requires the calculation of heating, ionization, and excitation by gamma-rays, X-rays, and diffuse UV/optical radiation in a multi-zone supernova ejecta structure taken from stellar evolution/explosion models, including NLTE solutions for the important atoms and a detailed radiative transfer treatment[19]. Such an analysis is beyond the scope here, and we instead aim to explore the range of densities and temperatures that can roughly reproduce the luminosities of the most prominent lines in the SN2007bi spectrum at 367d after peak. We compute NLTE solutions for oxygen, magnesium and iron, as functions of density and temperature, taking

---

[ix] https://c3.lbl.gov/es/

thermal processes and recombination into account. In these models, the variable parameters are:

$n_{\{OI, MgI, FeII\}}$ - Number density of the element (O I, Mg I, and Fe II, respectively);
$n_e$ - Number density of electrons;
$T$ - Temperature.

The fixed parameters are:

Velocity gradient (we assume homologous expansion: $dv/dr=1/t$);
$n_{OII} = 0.9\, n_e$ ;
$n_{MgII} = 0.1\, n_e$ ;
$n_{FeIII} = 0$,

where we assume that oxygen and magnesium dominate the composition of the ejecta in the region where oxygen/magnesium lines are produced. The 0.9:0.1 partition roughly reflects the relative abundances of oxygen and magnesium in stellar evolution models, where the ratio is close to the solar ratio of $n_O/n_{Mg} \sim 13$ (refs 52, 53), and that iron is singly ionized (recombinations are usually not important for the [Fe II] lines anyway).

We explored solutions in the range $10^0 < \{n_e, n_i\} < 10^{10}$ cm$^{-3}$ and for three temperatures: $T = 2000$, 5000 and 8000 K.

It can be seen from Extended Data Fig. 5 that the electron density must be close to $n_e = 10^7$ cm$^{-3}$ in order to reproduce the O I 7774 Å luminosity, which is a density-sensitive recombination line. At a temperature of 5000 K, the O I and Mg I densities needed to reproduce [O I] 6300, 6364 Å and Mg I] 4571 Å are $n_{OI} \sim 10^6$ and $n_{MgI} \sim 10^3$ cm$^{-3}$, respectively. That magnesium is more strongly ionized than oxygen is consistent with its much lower ionization potential.

This electron density can be checked for consistency against the mass we derive from the light curve (see Main Text and Section 5 below). If we assume single ionization ($n_{ion} \sim n_e$) then the total number of ions is $N_{ion} \sim n_e V$, where $V$ is the volume. The volume comes simply from spherical expansion at 10000 km s$^{-1}$ for 400 days, where these numbers are the FWHM of the Mg I] 4571 Å line and the approximate time since explosion (367d + rise time). Assuming oxygen dominated ejecta ($\bar{A} \sim 16$), we estimate an approximate mass:

$$M_{ejecta} \sim N_{ion}\, \bar{A}\, m_{proton} \sim 22\, M_\odot$$

Thus it is consistent with the mass derived from the light curve.

We then investigated what conditions are required to produce a strong [Fe II] 5200 Å line. We find that for $T < 3000$ K, the strongest [Fe II] line is 7155 Å (in agreement with previous models[1]), which is not observed. For temperatures

$T$ > 6000 K, [Fe II] 4330 Å will be the strongest visible line, which is not observed either. Between 3000-6000 K, the 5200 Å line (which is observed) should be the dominant feature. We therefore find that the temperature is likely to be in this intermediate regime. However, even in this regime the other [Fe II] lines remain at levels of 60-90% of the flux of the feature at 5200 Å, much higher than their observational limits. We therefore also conclude that Fe II can at most contribute only part of the 5200 Å feature.

Given the strong Mg I] 4571 Å line in the spectrum, one good candidate for contributing to emission around 5200 A is Mg I 5183 Å, which is the second strongest recombination line after Mg I] 4571 Å. This line is an allowed triplet (5167.32, 5172.68 and 5183.60 Å), with the latter two transitions dominating. As Extended Data Fig. 5 shows, the flux in this line can become a significant fraction of the Mg I] 4571 Å flux. In the regime where it is strong, the electron density is constrained to the same value as that from the O I 7774 Å recombination line, $n_e \sim 10^7$ cm$^{-3}$.

For a given electron density, we can use the temperature constraints on the iron-emitting zone to put an upper limit to the Fe II density. Using the range 3000 K < $T$ < 6000 K, and assuming a similar electron density to that derived for the oxygen/magnesium zones ($n_e \sim 10^6$-$10^8$ cm$^{-3}$), we find $n_{FeII} < 10^6$ cm$^{-3}$, and an Fe II mass of 0.001-1 M$_\odot$ (at 6000 and 3000 K, respectively). While a lower electron density in this zone would allow for a larger iron mass, this calculation shows that there are density regimes where small iron masses (and therefore small/moderate amounts of $^{56}$Ni) can reproduce the 5200 Å feature.

We conclude that a 10-20 M$_\odot$ core-collapse ejecta, dominated by oxygen/magnesium and <<1 M$_\odot$ of $^{56}$Ni, can reproduce the main lines observed in SN2007bi, given that some power source keeps the ejecta ionized (1-$x_e$ << 1) and hot (T ≳ 5000 K) for several hundred days. Energy input by an energetic (fast-spinning) and medium-fast decaying magnetar (spin-down time scale of months/years) is an excellent candidate for providing such physical conditions.

In summary, previous nebular modelling[1] of SN2007bi can explain the observed lines if the heating of a large ejecta mass is caused by radioactive $^{56}$Ni. However our calculations indicate that this is not the only physical scenario that can reproduce the nebular spectrum. Moreover, we find that the lack of an observed [Fe II] 7155 Å line, which should be detectable at the low temperatures and high iron mass in a PISN (and was also predicted by previous nebular models[1]), is problematic for a pair-instability interpretation of SN2007bi. We have shown here that the strong line at 5200 Å, previously interpreted as a 'smoking gun' signature of a very high-nickel-mass event, can be reproduced with small-to-moderate amounts of nickel under certain physical conditions (higher temperature, and some Mg I blending).

## 5 Bolometric light curve and alternative model fits

We derived a bolometric light curve for PTF12dam by converting magnitudes in near-UV, optical and NIR filters into physical fluxes, correcting for the extinction described in Section 3. We derive an SED at each epoch by linearly interpolating the flux between the effective filter wavelengths. The total flux was then converted to a luminosity using the distance derived from the redshift and our assumed cosmology (Section 2.1). This was done at every epoch with an *r* band observation, and magnitudes in other filters were interpolated to these epochs using low-order polynomials. Zero flux was assumed outside of the observed wavelength range (1700-23000 Å). Between epochs -11d and +21d, we have full restframe flux coverage from the UV to the NIR. Outside of this period, we make simple extrapolations to account for missing UV and NIR data. To correct the early epochs (i.e. more than 11d before peak) for missing UV and NIR data, we extrapolated the flux contribution by assuming a linear gradient in colour evolution in each filter with respect to *r*. This is reasonable, since the colour evolution is very close to linear over the epochs where we have full coverage. To correct the late epochs for missing UV data, we simply extrapolated the UV light curves linearly. This is because the rapid fall off in the near-UV after maximum light means that it contributes little to the total luminosity beyond ~ 40d post-peak, so the lack of coverage does not have a significant effect.

To determine the errors due to the missing near-UV coverage at early and late times, we make two bolometric light curves. Method 1 is the one described above. For method 2, we integrate only the observed bands at each epoch, and assume the fraction of flux in the UV at all epochs when it is no longer observed is the same as the fraction at -20d (early times) or +25d (late times). At these later epochs, the integrated luminosity is converted to a total luminosity by adding this missing fraction. The additional error in log(*L*) due to missing UV points is taken to be the difference in log(*L*) given by these two methods, and is included on Figure 4.

### 5.1 Light curve models powered by radioactive $^{56}$Ni

We modelled the total luminosity of PTF12dam using a semi-analytic treatment based on the Arnett diffusion solution for a specified power source in a radiation-dominated, homologously expanding ejecta[21,22]. The treatment is identical to our magnetar fit (see Main Text and Supplementary Information section 5.2), but with the magnetar power source replaced by $^{56}$Ni and $^{56}$Co decay. For the $^{56}$Ni-powered models, we initially adopted an explosion energy of $10^{51}$ erg and computed the ejecta mass and $^{56}$Ni mass that produced the best fit to the bolometric light curve. The best fit was formally determined through a $\chi^2$ minimization. We use $\chi^2$ per degree of freedom, with 16 degrees of freedom (19 data points minus 3 free parameters: ejecta mass, $^{56}$Ni mass, and time). We were unable to produce a satisfactory fit (see Extended Data Fig. 6) to the shape of the light curve, particularly around peak, and found ejecta masses of ~15 $M_\odot$ with quite unphysical $^{56}$Ni masses of greater than 14 $M_\odot$. The parameters for this best fitting model are given in Extended Data Fig. 6. Formally, better fits can be obtained with higher explosion energies of $10^{52}$ and $10^{53}$ erg, but again very high $^{56}$Ni masses of 14 $M_\odot$ and 17 $M_\odot$ are required. While this results in $M_{Ni}/M_{ej}$ ratios that are unphysically high for an iron core-collapse supernova, for the most energetic explosions we find ratios that could be produced in thermonuclear events. However, the associated total ejecta mass is not compatible with any proposed progenitor of such an explosion. $M_{ej} \lesssim 50$ $M_\odot$ is much too high for a SN Ia-like model. As for pair-instability models: producing more than 10 $M_\odot$ of $^{56}$Ni seems to require[7] helium cores of more than 110 $M_\odot$, so we would expect much more massive ejecta compared to the 50 $M_\odot$ in our fit. Thus, we are in a region of parameter space that does not correspond to any quantitative physical model. The same problem occurs in fitting similar models to a more typical SLSN, PS1-10bzj[54], with fits giving ejecta compositions that are >75% $^{56}$Ni. It should be noted that a good fit was found for PS1-10bzj with a magnetar-powered model similar to ours.

The core-collapse of a massive star (43 $M_\odot$ carbon-oxygen core from a 100 $M_\odot$ main-sequence progenitor) has also previously been proposed[10,20] to explain the SN2007bi light curve. A model[20] with an explosion energy (a free parameter) of $3.6 \times 10^{52}$ erg and $M_{ej} = 40$ $M_\odot$ reproduces the light curve with a $^{56}$Ni mass of $M_{Ni} = 6.1$ $M_\odot$. However this was based on the SN2007bi peak luminosity of $5.8 \times 10^{43}$ erg s$^{-1}$. PTF12dam is intrinsically brighter, and a full bolometric luminosity calculated from our UV to NIR coverage provides a measured peak luminosity of $1.2 \times 10^{44}$ erg s$^{-1}$. This difference, a factor of 2.1, explains why we require a much larger $^{56}$Ni to total ejecta mass ratio than that proposed[20] for SN2007bi. The measured luminosity of PTF12dam effectively makes core-collapse models unphysical due to the large $^{56}$Ni mass required. Even extreme massive-core-collapse models[20] produce only 4 $M_\odot$ of $^{56}$Ni, and none has $M_{Ni}/M_{ej} > 0.2$. Additionally, such models are only likely to be possible in extremely low metallicity environments of $Z \sim Z_\odot/200$. We therefore cannot find a physically reasonable fit to our light curve from this radioactive diffusion model.

### 5.2 Unexplored parameters in PISN models

The long rise times of existing PISN models[7,8] are central to our conclusion that the objects we observe are not pair-instability explosions, so it is important to establish how fundamental the slow rise is. Could more sophisticated models (taking into account rotation, magnetic fields, mixing and higher dimensions) have substantially shorter rise times?

Simple arguments show that they cannot. The rise time of a supernova is set by the diffusion timescale, given by[22]

$$t_{\text{diff}} = 440\text{d} \left(\frac{E}{10^{51}\text{erg}}\right)^{-1/4} \left(\frac{M_{\text{ej}}}{100\,M_\odot}\right)^{3/4} \left(\frac{\kappa}{0.2\text{ cm}^2\text{g}^{-1}}\right)^{1/2},$$

where $E$ is the kinetic energy of the explosion, $M_{\text{ej}}$ the ejected mass and $\kappa$ the opacity. In the rising phase, the gas will be highly ionized, so $\kappa$ must be close to 0.2 cm$^2$ g$^{-1}$. For a 100 $M_\odot$ sphere, this gives diffusion times of 246d or 140d for an explosion of energy $10^{52}$ or $10^{53}$ erg respectively. If the ejected mass of $^{56}$Ni is 5 $M_\odot$, a self-consistent set of parameters[7,8] is $E \sim 4\times10^{52}$ erg and $M_{\text{ej}} \sim$ 100 $M_\odot$. The diffusion time is then 175d.

Testing this calculation against our light curve fits (Extended Data Fig. 6), we find that $t_{\text{rise}} \sim 1.1\ t_{\text{diff}}$, confirming the validity of this approximation. The timescales calculated in this way are independent of more complicated effects like rotation or magnetic fields, as this is just radiative diffusion (in the particularly simple regime where electron scattering dominates the opacity). Even if we conservatively set the rise time as half of the diffusion time, a super-luminous pair-instability event like that proposed for SN2007bi should have $t_{\text{rise}} \gtrsim 100$d.

Multi-dimensional simulations of PISN explosions[55] have suggested that mixing in the ejecta is negligible. However, consequences of mixing have been studied for $^{56}$Ni-powered SNe Ia[56] and Ibc[57,58]. In fact, mixing serves to flatten the early parts of model light curves, since photons begin diffusing out earlier, without significantly decreasing the time to reach maximum light, as the bulk of the nickel still resides in the inner ejecta. Such models also display redder spectra[57]. These two effects mean that the one-dimensional models we find to be incompatible with PTF12dam and PS1-11ap are likely better fits than PISN models with detailed mixing would be.

### 5.3 Pulsational pair-instability supernovae

Another mechanism proposed to explain super-luminous supernovae, closely related to the PISN scenario, is the so-called 'pulsational pair-instability'. Stars with zero-age main sequence masses of 90-130 $M_\odot$ are expected to eject shells of material in a series of pair-instability-powered eruptions; collisions between successive shells can produce a bright display[59]. However, these interaction-driven events involve large amounts of hydrogen and/or helium[59,60] at relatively low velocities ($\sim 10^3$ km s$^{-1}$) and are therefore expected to display narrow H/He lines. Furthermore, to reach the luminosities of SLSNe ($L \sim 10^{44}$ erg s$^{-1}$), the first shell must have reached a significant radius before the second shell collides with it (in order to avoid adiabatic losses), and at typical radii of $10^{15}$-$10^{16}$ cm, the spectra produced are faint in the U and B bands[59]. The spectra of pulsational-PISNe must therefore be very different, exhibiting a low continuum temperature and narrow H/He lines, from those observed in our objects.

### 5.4 Details of the magnetar powered model

Our magnetar[16,17,61] model[14,21,22] assumes a SN explosion energy of $10^{51}$ erg and derives a magnetar luminosity from magnetic field and spin period; these are free parameters to fit. Assuming a 45° angle between spin axis and magnetic field (this can be fixed, as it simply serves to change the effective *B*-field), we feed the time-averaged magnetar luminosity into the Arnett diffusion solution[21,22], in the same way as for the radioactive model of section 5.1. The resultant luminosity of the SN is calculated, and the excess input energy goes into kinetic energy of the explosion. These models have been tested against more detailed simulations (via private communication with Dan Kasen, based on published models[14]) and found to yield good agreement[21]. The magnetar model can generate a wide range of light curves because the ejecta mass and power source are decoupled, such that we can find an ejecta mass to fit the observed diffusion timescale (along with explosion energy and opacity[22]), and an energy input to power the observed luminosity. We found that this model could fit the data with physically reasonable parameters for the explosion, ejecta and magnetar. The power source in a SLSN (be it a magnetar or something else) keeps the ejecta ionized for much longer than in a typical supernova, such that electron scattering provides a high continuum opacity. The opacity for a highly ionized, hydrogen-free gas is $\kappa \sim 0.1$-$0.2$ cm$^2$ g$^{-1}$. Within these limits, we find ejecta masses in the range 10-16 M$_\odot$.

This suggests a fairly massive progenitor, in which case we might expect black hole formation to be a more likely outcome. A central-engine model could still apply in such a scenario; however, the engine would then be accretion onto the black hole. Fallback models[62] predict an energy input similar to the magnetar, but with an asymptotic time dependence of $L \propto t^{-5/3}$ rather than $t^{-2}$.

We can check for consistency between our light curve fitting and observed spectral evolution by estimating the time taken for the ejecta to become nebular. We set the continuum optical depth, $\tau_c$, to unity in the expression

$$\tau_c = \kappa \rho v t$$

where $\kappa$ is the opacity, $\rho$ the density and $v$ the velocity of the ejecta, and $t$ is the time since explosion. For electron scattering, $\kappa = 0.2$ cm$^2$ g$^{-1}$, giving

$$t_{neb} \approx 360 \, d \left(\frac{M}{10 \, M_\odot}\right)^{1/2} \left(\frac{v}{10^4 \, km \, s^{-1}}\right)^{-1}.$$

Thus, we would expect the transition to the nebular phase to occur at approximately a year after explosion. PTF12dam is not yet fully nebular at ~280 d after explosion (221d post-peak spectrum, with 50-60d rise), while the transition in SN2007bi occurred somewhere between 134 and 367d after peak (all in respective rest frames).

We treat the magnetar input as remaining fully trapped for the 200 post-peak days of light curve data. Models predict[63] that most of the magnetar emission is initially transformed to kinetic energy of relativistic particles; the opacity to these is much higher than the optical opacity, and this wind will remain fully trapped at the base of the ejecta for a long time. However, much of the wind energy is converted to X-rays and gamma rays at the interface between the wind and the ejecta, so opacity at these wavelengths becomes important. Calculations of the emission and reabsorption of this radiation is hampered by several uncertain physical processes operating at the interface, and how partitioning between magnetic, thermal and non-thermal electron and ion pools occur[63]. For a 3 ms pulsar (with a magnetic field of $10^{13}$ G) in 5 $M_\odot$ of hydrogen- and helium-dominated ejecta, gamma rays start to escape at a significant rate after ~100 d (ref. 63). For PTF12dam, we expect a longer timescale, as the higher ejecta mass (15 $M_\odot$ rather than 5 $M_\odot$) and a greater opacity (metal- rather than hydrogen-dominated) both serve to increase gamma ray trapping. Thus, our assumption of full trapping over the observed period is a reasonable one.

## 5.5 Temperature evolution

Spectral models[13] show that one of the key observables to distinguish between magnetar-powered super-luminous supernovae and PISNe is the colour of the spectra: the extremely massive progenitors of PISNe result in low energy-to-ejecta-mass ratios and therefore their spectra should be much cooler. We estimated effective temperatures for PTF12dam and SN2007bi using two methods. We fitted blackbody curves to the optical photometric flux, and then to the continuum in the spectra. The two methods gave similar temperature estimates. Our magnetar light curve model also provides the photospheric temperature evolution. The radius of the ejecta is derived from the input kinetic energy and elapsed time; the effective temperature can then be estimated from the luminosity and radius using the Stefan-Boltzmann law.

We calculate an effective temperature for the PISN models[7] by fitting a blackbody curve to synthetic photometry derived from these model spectra. The results are shown in Extended Data Fig. 4. We can see that the magnetar model better matches the observed evolution: with high early temperatures followed by a steep decline before flattening, though our observed rise seems to be slower than the simple model predicts. We note, however, that the early temperatures are not so well constrained, as we have only photometry and no spectra at these epochs. The PISN models show a relatively constant temperature phase of ~100d before declining just after maximum light, as expected from the increase in IGE line blanketing, and do not reach temperatures above $10^4$ K, in contrast to our observations.

# 6  Do PISNe exist?

## 6.1 PISN rate within $z < 0.6$

Our Monte-Carlo simulation[25] generates a random array of redshifts (within the range $0 < z < 1$) and explosion epochs within the survey year for 50000 objects based on PISN models[7], and computes observer frame light curves including *k*-corrections, time dilation and average extinction. The rate of occurrence as a function of redshift is computed from the change in volume and cosmic star-formation history, assuming that a fixed fraction (0.007) of the star formation leads to core-collapse, and that the PISN rate is itself a fixed fraction of the core-collapse rate. The models are then subjected to a simulated survey with the depth and cadence of the PS1 Medium Deep Survey in order to calculate the number we should expect to see.

As a sanity check, we also carry out a rough manual calculation of the approximate rate limit we would expect to recover. The PS1 Medium Deep survey fields cover 70 square degrees[23], which corresponds to a volume within redshift $z < 0.6$ of 0.07 Gpc$^3$. The brightest models have absolute peak magnitudes in the NUV of $M_U \sim -22$ (AB magnitude; refs 12,24), giving apparent peak magnitudes $r_{P1}$, $i_{P1} \lesssim$ 20-21. The typical PS1 nightly detection limit in these filters is 23.5 (refs 21, 23, 24), which would allow them to be detected 2-3 magnitudes before and after maximum light (130-200d in observer frame). The observing window for each Medium Deep field during a single year is 150d. We assume that the survey is sensitive to about 50% of all the $z < 0.6$ PISNe reaching these magnitudes at maximum light during this window. They would be readily identifiable by their high luminosity and slow rise and decline over the season, but we may miss the detection of a peak in around 50% (such events would have incomplete lightcurves). In 3 years of the Medium Deep Survey, we have detected no supernova-like transients with these peak magnitudes exhibiting a PISN-like rise time (within $z < 0.6$), and only PS1-11ap has shown a slow decline (e.g. refs 2, 24, 64). We thus conclude that the super-luminous PISN rate must be less than 10 Gpc$^{-3}$ yr$^{-1}$, or, using the core-collapse SN rate at $0.5 < z < 0.9$, (ref. 26; $4\times10^{-4}$ SNe Mpc$^{-3}$ yr$^{-1}$), this corresponds to less than $\sim 10^{-5}$ of the core collapse rate, in agreement with the results of our simulation. These bright PISN candidates thus appear to be a factor of a few (perhaps up to 10) less common than the general population of SLSNe of type I and type Ic[1,12].

## 6.2 PISN candidates at higher redshift

Two SLSN, at redshifts 2.05 and 3.90, were recently presented[4], and suggested to be possible pair instability supernovae. These objects display rise times similar to PTF12dam, though not so well constrained, and are shown to fit PISN model light curves[7] quite well (the same models that PTF12dam clearly does not match; see Figs 1 and 4). This apparent contradiction results from their high redshifts. The observed optical photometry actually probes restframe UV. The PISN models quickly fade in the UV relative to the optical, due to line blocking[7,8]. Integrating the SEDs of these models between 1500 and 2500 Å, the approximate wavelength range sampled by these observations[4], we confirm the fit: the resultant light curves do indeed reach peak in ~50 days, before fading to negligible brightness in a further ~150 days. Thus, we cannot exclude the possibility that these high-redshift SLSNe are the first observed examples of PISNe. However, the lack of spectra and restframe optical data preclude a firm classification.

PISNe were originally expected only at high redshift, as the required massive cores are difficult to form except at low metallicities. While we still expect very massive stars and PISNe to be more common in the early Universe, it should be noted that recent simulations[65] show that stars of initial mass 100-290 $M_\odot$ may end their lives in hydrogen-free PISNe at SMC metallicity.